\documentclass[reprint,eqsecnum,floats,aps,amsmath,amssymb,nofootinbib,prd,showpacs]{revtex4-1}

\usepackage[utf8]{inputenc}
\usepackage{amsfonts,amssymb}
\usepackage{amsmath}
\usepackage{hyperref}
\usepackage{graphicx}
\usepackage{subfigure}
\usepackage{enumerate}

\newcommand{\id}{\openone}
\newcommand{\re}{\mathbb{R}}
\newcommand{\rd}{{\rm d}}

\newcommand{\Hil}{\mathcal{H}}

\newcommand{\diff}{{\rm diff}}

\newcommand{\BohrOne}{{L^2(\bar{\re}_{{\rm Bohr}},\rd\mu_{{\rm Bohr}})}}

\usepackage[active]{srcltx}

\allowdisplaybreaks

\begin{document}

\title{Loop quantization of the Gowdy model with local rotational symmetry}

\author{Daniel Mart\'{\i}n de Blas}
	\email{damartind@uc.cl} 
	\affiliation{Instituto de F\'isica, Pontificia Universidad Cat\'olica de Chile, Avenida Vicu\~na Mackenna 4860, Santiago, Chile}
	\affiliation{Instituto de Estructura de la Materia, IEM-CSIC, Serrano 121, 28006 Madrid, Spain}
\author{Javier Olmedo}
	\email{jao44@psu.edu}
	\affiliation{Instituto de F\'isica, Facultad de Ciencias, Igu\'a 4225, esq. Mataojo, 11400 Montevideo, Uruguay} 
	\affiliation{Department of Physics and Astronomy, Louisiana State University, Baton Rouge, LA 70803-4001, USA}
	\affiliation{Institute for Gravitation and the Cosmos \& Physics Department, The Pennsylvania State University, University Park, PA 16802 U.S.A.}
\author{Tomasz Paw{\l}owski}
	\email{tpawlowski@cft.edu.pl} 
	\affiliation{Center for Theoretical Physics, Polish Academy of Sciences, Al. Lotnik\'ow 32/46, 02-668 Warsaw, Poland.}
	\affiliation{Instytut Fizyki Teoretycznej, Uniwersytet Warszawski, Pasteura 5, 02-093 Warszawa, Poland, EU.}
	
\begin{abstract}
	We provide a full quantization of the vacuum Gowdy model with local rotational symmetry. We consider a redefinition of the constraints where the Hamiltonian Poisson-commutes with itself. We then apply the canonical quantization program of loop quantum gravity within an improved dynamics scheme. We identify the exact solutions of the constraints and the physical observables, and we construct the physical Hilbert space. It is remarkable that quantum spacetimes are free of singularities. New quantum observables naturally arising in the treatment partially codify the discretization of the geometry. The preliminary analysis of the asymptotic future/past of the evolution indicates that the existing Abelianization technique needs further refinement.
\end{abstract}

\maketitle

\section{Introduction}

Making realistic predictions on effects of quantum gravity in the cosmological context---an element needed in particular to solve the singularity problem in cosmology---requires investigating models admitting inhomogeneous spacetimes, preferably at the nonperturbative level. Among such settings, Gowdy spacetimes \cite{gowdy1} in vacuum are particularly interesting since they are a natural extension of Bianchi I cosmologies \cite{bianchi} and admit nonperturbative inhomogeneities in the form of gravitational waves.

As they capture essential properties of full general relativity (GR) and at the same time are relatively simple, these models have brought over the years a lot of attention of researchers delving upon various aspects of gravity quantization. For instance, a quantum (geometrodynamics) description was already considered in the 1970s \cite{misner,berger1}.
Further they were explored in the specific context of gravitational waves quantization \cite{pierri} which description was later shown to admit a unitary dynamics \cite{unit1,*unit2,*unit3,*unit4,*unit5}. Besides, it was possible to prove that this representation where the dynamics is unitarily implementable is indeed unique \cite{uniq1,*uniq2} if in addition it is compatible with the symmetries of the equation of motion, i.e. the symmetries of the dynamics. The canonical quantization of these models employing the (complex) Ashtekar variables was carried out in Ref.  \cite{mena1}. These studies however, as treating the homogeneous background either classically or via geometrodynamics could not 'cure' the singularity problem.

The Gowdy model with linear polarization and $T^3$ spatial slices has been subsequently studied in terms of real Ashtekar--Barbero variables \cite{baren,baren2} via the midisuperspaces techniques \cite{bs-spher}. There, however, the difficulties in applying the conventional loop quantum gravity (LQG) techniques \cite{qsd} did not allow 
to complete the quantization program and probe the dynamics. Fortunately, the problems hampering prior approaches were successfully addressed in Ref. \cite{hybrid,hybrid1,*hybrid2} via the so-called \emph{hybrid quantization} program. This approach, also suitable for perturbative cosmological scenarios \cite{hyb-inf,hyb-inf1,*hyb-inf2,*hyb-inf3,*hyb-inf4,*hyb-inf5,*hyb-inf6}, combines the standard Fock quantization for the gravitational waves with a polymeric quantization of the homogeneous degrees of freedom. 
It is furthermore quite convenient for the study of quantum gravity in the presence of matter \cite{hybrid3}, and provides an arena for unveiling novel quantum phenomena on some sectors of the theory \cite{hybrid4,*hybrid5,*hybrid6}. All these models allow for a convenient partial gauge choice which reduces the set of local constraints to a global Hamiltonian and diffeomorphism constraints. The latter are sufficiently simple to allow finding their solutions (at least formally) and to construct the physical Hilbert space. This approach however, while successful, by the very nature of the hybrid quantization cannot be easily related with the standard LQG.

In this work we follow a more orthodox approach, expanding upon the original midisuperspace program of Ref. \cite{mop-lrs}. In order to test new techniques we study a slight simplification of the full polarized, three-torus Gowdy model, namely its locally rotational symmetric (LRS) version, where one identifies the two directions orthogonal to the inhomogeneous one. 
While in vacuo the model, being diffeomorphic to a homogeneous Bianchi I spacetime (Kasner solutions), features just one free global degree of freedom, in presence of matter (for example a massless scalar field) it admits genuine inhomogeneous solutions, containing however a homogeneous and even isotropic sector.
This feature makes the model viable for cosmology applications and particularly useful in testing the results of the perturbative approaches against nonperturbative effects as well as in the studies of the relation of loop quantum cosmology (LQC) \cite{lqc} with LQG. For instance, Bianchi I spacetimes in loop quantum cosmology \cite{awe-b1,mmp-B1-evo}, after imposing local rotational symmetry, or the hybrid quantization of the polarized Gowdy model in the three-torus \cite{hybrid,hybrid1,hybrid2,hybrid3} are clearly interesting for this purpose. Gowdy LRS has also been consider in the context of the (loop) consistent algebra approach \cite{bojo-bra}. Our quantization program, unlike the previous approaches, will not involve gauge-fixing. Instead, we will be working with the constraint algebra, featuring (as in full GR) \emph{local} structure functions, and will be forced to employ the Dirac program in a manner featuring the same level of complication as in full LQG. To deal with the known difficulties in its implementation we will follow a strategy already adopted in studies of spherically symmetric spacetimes \cite{BH-1,BH-2} (see Ref. \cite{bb-gowdy} for a discussion of the full polarized Gowdy model). That strategy is based on a specific redefinition of the constraints and consequently of their algebra structure, which makes the Hamiltonian constraint Abelian. Furthermore, in the construction of the quantum counterpart of this constraint we implement, for the first time in a loop quantized inhomogeneous model, an improved dynamics scheme. The solutions to the Hamiltonian constraint can be explicitly determined and can be equipped with a well defined Hilbert space structure, which in turn, together with the application of the standard LQC treatment of the spatial diffeomorphisms, allows us to unambiguously probe the dynamical sector of the model. It is remarkable that the resulting spacetimes are free of singularities. Furthermore, the area of the Killing orbits is quantized due to the discreteness of the spectrum of a new observable emerging in this quantization. This treatment and its results open a new window for the quantization of cosmological scenarios in LQC featuring a natural connection with the full theory by means of more realistic models admitting non-perturbative inhomogeneities.

The paper is organized as follows. In Sec. \ref{sec:class} we introduce the classical polarized Gowdy model on the three-torus. Then we consider the model with local rotational symmetry in Sec. \ref{sec:classLRS}, where we also provide the classical Dirac observables of the model. In Sec.~\ref{sec:abelian} we define the Abelian constraint. In Sec. \ref{sec:kin} we describe the kinematical quantum framework and the basic kinematical observables, whereas the Hamiltonian constraint is discussed in Sec. \ref{sec:qham}. The physical Hilbert space as well as the observables (together with a discussion about the semiclassical sectors of the theory) are provided in Sec. \ref{sec:phys}. We conclude with Sec. \ref{sec:conc}. Furthermore, in App. \ref{app:altern} we discuss an alternative construction for the Abelianized Hamiltonian constraint, and in App. \ref{app:disp-rel} a partial spectral analysis of some operators relevant for our treatment.

\section{Classical Polarized Gowdy model in Ashtekar-Barbero variables}
\label{sec:class}

Let us start by summarizing the description of the Gowdy model with spatial sections isomorphic to the three-torus and linear polarization for the gravitational wave content. We will provide this description in terms of real Ashtekar-Barbero variables with the notation introduced in Ref.~\cite{baren}. 
The polarized Gowdy model in this case consists of three local degrees of freedom $({\cal A},{\cal E})$\footnote{The connection ${\cal A}$ in our work is decreased by a factor of $2$ with respect to the one introduced in \cite{baren}.}, $(K_x, E^x)$ and $(K_y, E^y)$, all of them with support on the circle of angular coordinate $\theta\in[0,2\pi)$. They are pairs of connections and densitized triads, respectively, such that the spatial metric components are $g_{\theta\theta}=(E^{x}E^{y})/\mathcal{E}$, $g_{xx}=E^y\mathcal{E}/E^x$ and $g_{yy}=E^x\mathcal{E}/E^y$, and where the function $\mathcal{E}$ corresponds to the area of the (here 2-dimensional) orbits of the spatial Killing vectors. 

The symplectic structure is given by
\begin{equation}\label{eq:G-sympl}
	\Omega = \frac{1}{\gamma\kappa}
	\int{\rm d}\theta(2{\rm d}{\cal A}\wedge{\rm d}{\cal E}+{\rm d}K_x\wedge{\rm d}E^x+{\rm d}K_y\wedge{\rm d}E^y),
\end{equation} 
with\footnote{In order to restore the Immirzi parameter \cite{immirzi} in our studies we drop the rescaling by $\gamma$ introduced in \cite{baren}.}
\begin{equation}
\kappa := \frac{8\pi G}{4\pi^2}.
\end{equation}
Besides, the Hamiltonian of the model is a linear combination of two first class constraints: the diffeomorphism one
\begin{equation} 
C  =  \frac{1}{\gamma\kappa}\left[ (\partial_{\theta}K_x)E^x +
(\partial_{\theta}K_y)E^y - 2(\partial_{\theta}\cal{E})\cal{A} \right],
\label{FinalVector}
\end{equation}
and the Hamiltonian constraint
\begin{equation}\begin{split} 
	H & = \frac{1}{\kappa} \left[ - \frac{1}{\gamma^2\sqrt{E}} \bigg\{(K_x E^x K_y E^y)
		+ 2(K_x E^x +  K_y E^y)\cal{E} \cal{A}\bigg\} \right. \\ 
	& \left. - \frac{1}{4 \sqrt{E}}\bigg\{
		(\partial_{\theta}{\cal{E}})^2
		-\left({\cal{E}}\partial_\theta(\mbox{ln}(E^y/E^x))\right)^2 \bigg\} +
		\partial_\theta\left(\frac{\cal{E}\partial_\theta \cal{E} }{\sqrt{E}}\right)
	\right], \label{FinalHam}
\end{split}\end{equation}
where $E={\cal E}E^x E^y$. This set of constraints has a nontrivial algebra
\begin{subequations}\label{eq:G-algebra}\begin{align}
	\left\{C[N^{\theta}], C[M^{\theta}]\right\} & =  C\left[
		N^{\theta}\partial_{\theta}M^{\theta} -
		M^{\theta}\partial_{\theta}N^{\theta}\right] \\
	\left\{C[N^{\theta}], H[N]\right\} & = 
		H[N^{\theta}\partial_{\theta} N] \\
	\left\{H[M^{\vphantom{\theta}}], H[N]\right\} & =  C\left[\frac{{\cal E}^2(M\partial_{\theta}N -
		N\partial_{\theta}M)}{E} \right] .
\end{align}\end{subequations}
 
As we see, the constraint algebra involves structure functions, as it happens in the full theory. This is one of the main handicaps preventing us from attaining a complete quantization of the full polarized Gowdy model so far. One of the strategies that can be adopted in this type of field theories is to replace the original form of the Hamiltonian constraint with its Abelian version (see \cite{bb-gowdy} for the discussion of this method and its limitations). This procedure was originally suggested and successfully applied in spherically symmetric spacetimes \cite{BH-1,BH-2}. However, at the quantum level, the anomalies that a loop quantization introduces are not yet well understood in the context of the full polarized Gowdy model \cite{bb-gowdy}, thus preventing the completion of the traditional loop quantization program there.
Fortunately, as we will show further in the manuscript, the Abelianized version of the Hamiltonian constraint can be easily represented at the quantum level as an operator with a constraint algebra free of anomalies provided that we introduce an additional symmetry: \emph{local rotational symmetry}.

\section{Classical model with local rotational symmetry}
\label{sec:classLRS}

Let us consider a restriction of the standard polarized Gowdy model via imposing the requirement that the rotations on the $x-y$ plane are isometries. An easy way to represent the restricted geometries is to identify the degrees of freedom represented by the pairs $(K_x, E^x)$ and $(K_y, E^y)$. The original symplectic structure \eqref{eq:G-sympl} reduces then to
\begin{equation}
	\Omega=\frac{2}{\gamma\kappa}\int{\rm d}\theta({\rm d}{\cal A}\wedge{\rm d}{\cal E}+{\rm d}K_x\wedge{\rm d}E^x).
\end{equation} 
The phase space is now coordinatized by two pairs of canonical variables $({K}_x,E^x)$ and $(\cal A,\cal E)$. The spatial metric components become then
\begin{equation}
	g_{\theta\theta}=(E^{x})^2\mathcal{E}^{-1}, \qquad g_{xx}=g_{yy}=\mathcal{E}.
\end{equation} 

In the specified notation\footnote{The reader must be concerned about the fact that we keep here the same notation for the Hamiltonian and the constraints than in the full polarized Gowdy model. In order to avoid any possible confusion, in what follows we will refer only to the scenario with local rotational symmetry.} the Hamiltonian is again a linear combination of the constraints $ H_T=\int d\theta (NH+N_\theta C)$, where 
\begin{subequations}\begin{align}
	\begin{split}
		H=& - \frac{1}{\kappa}\left[\frac{1}{\gamma^2\sqrt{{\cal E}}} (K_x^2 E^x)
		+ \frac{4}{\gamma^2\sqrt{{\cal E}}}{\cal E}{\cal A}K_x \right.\\
		&\left.+ \frac{1}{4 \sqrt{{\cal E}}E^x}
		(\partial_{\theta}{\cal{E}})^2 
		- \partial_\theta\left(\frac{\cal{E}\partial_\theta \cal{E} }{\sqrt{{\cal E}}E^x}\right)\right],
	\end{split} \\
	C=&\frac{2}{\gamma\kappa}\left[(\partial_{\theta}K_x)E^x - (\partial_{\theta}\cal{E})\cal{A}\right].
\end{align}\end{subequations}

The constraint algebra remains unchanged with respect to Eq. \eqref{eq:G-algebra}
\begin{subequations}\begin{align}
	\left\{C[N^{\theta}], C[M^{\theta}]\right\} & =  C\left[
		N^{\theta}\partial_{\theta}M^{\theta} -
		M^{\theta}\partial_{\theta}N^{\theta} \right] \\
	\left\{C[N^{\theta}], H[N]\right\} & = 
		H[N^{\theta}\partial_{\theta} N] \\
	\left\{H[M^{\vphantom{\theta}}], H[N]\right\} & =  C\left[\frac{{\cal E} (M\partial_{\theta}N -
		N\partial_{\theta}M) }{(E^x)^2} \right] . \label{eq:Poisson-HH}
\end{align}\end{subequations}

In this restricted setting we are still dealing with a 1+1 field theory with a nontrivial constraint algebra as in the full polarized Gowdy model, sharing with the latter the level of complication and set of difficulties in implementing the Dirac quantization procedure.
On the other hand, the structure of the degrees of freedom changes drastically with respect to the original Gowdy model, as the local rotational symmetry fixes an infinite number of physical degrees of freedom. On shell, instead of two global and one local degree of freedom the model we study features only one global and no local ones---the degrees of freedom of Bianchi I Kasner solutions.
While in itself it does not admit genuinely inhomogeneous spacetimes\footnote{Although the solutions represent Kasner spacetimes, these spacetimes are foliated by Cauchy surfaces which are not invariant with respect to finite symmetry transformations.}, it describes the geometry in a diffeomorphism invariant manner resembling the one of full LQG and not tied to the homogeneity of the physical solutions. Because of that, the treatment can be extended in a straightforward way to LRS Gowdy models including matter content, in particular a massless scalar field. The latter, while it is in general genuinely inhomogeneous, it admits homogeneous (Bianchi I) and isotropic (Friedmann--Robertson--Walker solutions) spacetimes. Hence, it is strongly relevant for the study of our Universe. 

Since we are dealing with a constrained system, in order to provide the classical description, one needs to perform one more step: the identification of Dirac observables. For as long as we consider just the vacuum solutions (which is the case of this article), they just admit one global degree of freedom, thus one needs to identify just a single observable. One of the straightforward choices follows from the treatment of spherically symmetric black holes \cite{BH-2} (of which techniques we implement in this work), where one could choose the phase space function encoding the ADM mass. Using its form in Ashtekar-Barbero variables we can propose its ``analog'' (just in form but not in the physical meaning) in the LRS Gowdy context. It is defined as
\begin{equation}\label{eq:h1-def}
	{\mathfrak h}_1= \int\rd\theta \left( \frac{2}{\gamma^2}\sqrt{{\cal E}}K_x^2-\frac{\sqrt{\cal E}(\partial_\theta{\cal E})^2}{2(E^x)^{2}} \right).
\end{equation}
By a quite lengthy but straightforward calculation one can show that its Poisson bracket with the total Hamiltonian
\begin{equation}\begin{split}
	\left\{{\mathfrak h}_1,H_T(N,N^{\theta})\right\} &= - \int\rd\theta N \kappa\frac{{\cal E}\partial_{\theta}{\cal E}}{(E^x)^3} C \\
	&- \int\rd\theta N^{\theta} \kappa\left( \frac{\partial_{\theta}{\cal E}}{E^x} H - \frac{2 K_x\sqrt{{\cal E}}}{\gamma E^x} C \right)  , 
\end{split}\end{equation}
thus it vanishes on shell.\footnote{We must notice that, as one can deduce from the analysis in Sec. \ref{sec:abelian}, the spatial derivative of the integrand in the right-hand side of Eq. \eqref{eq:h1-def} is already a combination of constraints. Therefore, only its homogeneous mode, given by ${\mathfrak h}_1$, is non-vanishing on shell.} As a consequence, ${\mathfrak h}_1$ encodes the diffeomorphism-invariant constant of motion.

The classical description constructed above could be used as the basis for the (loop) quantization. In its present form, however, due to the nontrivial Poisson bracket \eqref{eq:Poisson-HH}, the resulting quantum Hamiltonian constraint would not be commuting with itself, which poses certain challenges in the completion of the quantization program. Therefore, following \cite{BH-1,BH-2} we will perform the modification of the classical description known as \emph{Abelianization procedure}.

\section{Abelianization of the scalar constraint in the reduced model}\label{sec:abelian}

In this section we will consider a redefinition of the scalar constraint so that the analog of the Poisson bracket \eqref{eq:Poisson-HH} (with the redefined constraint) vanishes strongly.

It is well known \cite{kuchar,*teitel} that for diffeomorphism invariant systems, transformations like
\begin{enumerate}
	\item $H\to\Omega(Q)H$, where $\Omega$ is a function of the configuration variables only (here denoted as $Q$ for simplicity) and $\Omega(Q)\neq 0$,
	\item $C_a\to \Lambda_a^b(Q)C_b$ with $\det \Lambda\neq 0$,
	\item $H\to H+P^a\Lambda_a^b(Q)C_b$ with $P$ representing momentum variables (conjugate to $Q$) and $\det \Lambda\neq 0$,
\end{enumerate}
provide a new set of constraints with the same constraint surface. 
Of course, these transformations are not canonical transformations since the constraint algebra is changing (i.e. the Poisson brackets between constraints change). They can be understood as redefinitions of the lapse and shift functions. In our case we will consider the last type of transformations in order to achieve a new scalar constraint such that it commutes with itself, while the rest of the constraint algebra structure remains that of a proper Lie algebra (with structure constants instead of structure functions).

This method was developed (and successfully applied) in the context of spherically symmetric black hole spacetimes in \cite{BH-1,BH-2}. Following the construction presented there we consider the following transformation
\begin{equation}\label{eq:abel1}
	H \to H - 2K_x\sqrt{\cal E}(\gamma\partial_\theta{\cal E})^{-1}C ,  \qquad
	C \to C , 
\end{equation}
which is the unique available transformation removing the dependence on ${\cal A}$ from $H$ and not involving rescaling $H$.
This transformation is unfortunately singular---not defined on any phase space point on which $\partial_{\theta}{\cal E}$ vanishes on any point in space. Due to the compactness of the spatial slices this is the case for all the considered geometries ($\partial_{\theta}{\cal E}(\theta) = 0$ for at least two isolated values of $\theta$). By invoking certain completion constructions, one can still work with it at the cost of cutting off from the phase space all the geometries where $\partial_{\theta}{\cal E}(\theta) = 0$ on an open interval. We will discuss that treatment and its limitations in Appendix \ref{app:altern}.

To avoid cutting-off a physically relevant portion of the phase space, we propose another transformation supplementing Eq. \eqref{eq:abel1}
with a rescaling by $\partial_{\theta}{\cal E}(E^x)^{-1}$, that is
\begin{equation}\label{eq:abel2}
	H \to (E^x)^{-1} [ (\partial_{\theta}{\cal E})H - 2K_x\sqrt{\cal E} \gamma^{-1} C  ] , \quad
	C \to C , 
\end{equation}
Such transformation has been considered in the context of full polarized Gowdy model in Ref. \cite{bb-gowdy}. 
While the term $(E^x)^{-1}$ is regular outside of the strong classical singularity, $\partial_{\theta}{\cal E}$ still makes the transformation singular, this time enlarging the constraint surfaces. This is caused by the fact that  the new sets of constraints are satisfied automatically on homogeneous configurations (they vanish identically) while the old set of constraints selects a proper sub-surface (within the surface of the homogeneous geometries) of the phase space. However, the spatial continuity of the classical fields 
(here playing the role of the phase space variables) implies the equivalence between the two sets of constraints on a subset of geometries (points in the phase space) admitting an open region where $\partial_{\theta}{\cal E}\neq 0$,   
\footnote{By continuity the old Hamiltonian constraint has to be satisfied at the boundary of the set $\partial_{\theta}{\cal E}\neq 0$, then by the diffeomorphism constraint the interior of the complement of this set has to correspond to the fully homogeneous spatial regions and all the variables must remain constant (in particular the same as on the boundary) at each connected subset of that complement. Thus the old constraints have to be satisfied also in the complement.}
, thus the expansion is nontrivial on the surface corresponding to the \emph{homogeneous geometries only}.
Therefore, at this point, it is necessary to stress that we are modifying our classical theory, which thus will not fully coincide with GR in its globally homogeneous sector. In the classical theory such geometries are highly nongeneric. Unfortunately the discreteness of the variables introduced by the polymeric quantization will make the set of states corresponding to these geometries a non-zero measure one and a certain level of care will be needed when studying the properties of the physical states.

Under the new transformation the total Hamiltonian $ H_T=\int \rd\theta (\tilde N\tilde H+\tilde N_\theta C)$ takes the form
\begin{equation}\label{eq:H-tot-class}
	\begin{split}
		H_T =&\frac{1}{\kappa}\int \rd\theta\;
			\bigg\{{\tilde N}\partial_\theta\left[-\frac{2}{\gamma^2}\sqrt{{\cal E}}K_x^2+\frac{\sqrt{\cal E}(\partial_\theta{\cal E})^2}{2(E^x)^{2}}\right]\\
		&+\frac{2}{\gamma}\tilde N_\theta
		\left[ (\partial_{\theta}K_x)E^x - (\partial_{\theta}\cal{E})\cal{A} \right]\bigg\}.
	\end{split}
\end{equation}

The above redefinition of the constraint algebra does not affect the properties of the Dirac observable ${\mathfrak h}_1$ defined in Eq. \eqref{eq:h1-def}, which still remains a weak Dirac observable. Since the new Hamiltonian constraint is a linear combination of the original Hamiltonian and diffeomorphism constraints, one can check explicitly that ${\mathfrak h_1}$ commutes (under Poisson brackets) with the new set of constraints on shell. Actually, one can see that the integrand in the right-hand side of Eq. \eqref{eq:h1-def} is related  with the new scalar constraint---see the phase space function inside the square brackets in Eq. \eqref{eq:H-tot-class}.

\begin{figure}
\centering
\includegraphics[width=0.45\textwidth]{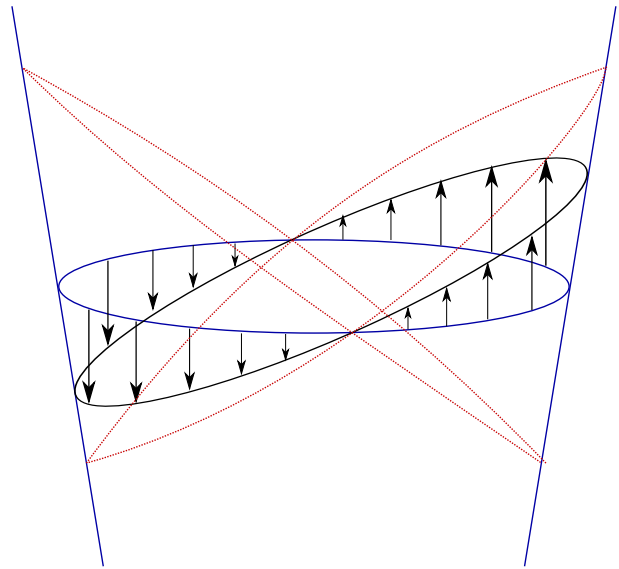}
\caption{Pictorial representation of the time evolution generated by the new scalar constraint. 
Vertical arrows represent time-like vectors. The blue circle is the initial Cauchy surface, the black circle 
is the evolved Cauchy surface to some time close to the initial one, and the red arcs represent 
the null boundaries of the region which can be reached by the Hamiltonian evolution generated by the new constraint.}
\label{fig:evol}
\end{figure}

At this point a few remarks on the range of the classical evolution generated by the new Hamiltonian are necessary. For that, let us compare the Hamiltonian flow generators or the $\partial_t$ vectors in both approaches. The original lapse function $N$ is related to the new one $\tilde N$ as follows 
\begin{equation}
  N = \tilde{N} \frac{\partial_{\theta}{\cal E}}{E^x} . 
\end{equation}
Since to be well defined the infinitesimal time translation vector does need to be finite, so does $\tilde{N}$. On the other hand the topology of the reduced manifold enforces the existence of at least two points on ${\cal S}$ where $\partial_{\theta}{\cal E}=0$. At each of these points the new time translation generators will necessarily vanish---will not generate the evolution in the original formulation of the theory as the reduction of GR. Furthermore, for the canonical formalism to work, the constant time slices have to remain spatial. As a consequence, the range of the classical evolution after Abelianization is severely limited, as the constant time slices must stay causally disconnected from any point at which $\partial_{\theta}{\cal E}$ vanishes. The time evolution determines the geometry only within spacetime regions containing the initial data slice and bounded by both the future and past light cones of the points (on the initial data surface) where $\partial_{\theta}{\cal E}=0$. In Fig. \ref{fig:evol} we provide a schematic diagram of the time evolution of the spacetime. 
As the attempt to increase the range of the evolution one can consider alternative Abelianization procedure, skipping for example the step \eqref{eq:abel2}. We discuss such approach in Appendix~\ref{app:altern}. Unfortunately, due to the singular nature of the transformation, also inherited in the quantum theory, we have decided to keep away of this alternative in the principal treatment of this manuscript.

In summary, in this section we obtained a complete classical description with a strongly self-commuting Hamiltonian constraint. This setting will be used in the remaining part of the article to build a (loop) quantum description of this LRS Gowdy model. The first step in this building process is the so-called \emph{kinematical level quantization}, where one ignores the constraints.

\section{Quantum kinematical theory}\label{sec:kin}

We will proceed now with the quantization of the above system within loop quantum gravity \cite{qsd}. It has been already specified in Refs. \cite{baren,baren2}, following the ideas of Ref. \cite{b-spher,bs-spher,bs-ham}.

Here the basic objects of the description are: 
\begin{itemize}
	\item $1$-dimensional closed oriented graphs embedded on the reduced manifold ${\cal B} := T^3/S$ (Where $S$ 
		are $2$-dimensional surfaces generated by isometries). Each graph contains a collection of disjoint
		edges $e_j$ terminating in vertices $v_j$.
	\item An algebra of holonomies along disjoint (oriented) edges of these graphs and the intertwiners on their vertices. 
	\item A basis of states $|\vec{k},\vec{\mu}\rangle$ which in the connection representation is given by
		\begin{equation}\label{eq:Hkin-base}
			\langle A|\vec{k},\vec{\mu}\rangle=\prod_{j}\exp\left\{i\frac{k_j}{2}\int_{e_j}\rd\theta {\cal A}(\theta)\right\}\exp\left\{i\frac{\mu_j}{2}K_x(\theta_j)\right\}.
		\end{equation}
		where $k_j\in \mathbb{Z}$ and $\mu_j\in \mathbb{R}$ are valences of edges\footnote{The signs of the (otherwise natural) valences 
			encode the orientation of a particular edge.} 
		$e_j$ and vertices $v_j$, respectively. These valences enumerate the representations of holonomy sub-algebras
		(correspondingly, holonomies along ${\cal B}$ and point holonomies encoding parallel transports along curves on $S$).
\end{itemize}

In this manuscript, we will adhere to the so-called {\it improved dynamics} scheme. It was originally proposed in the context of isotropic LQC \cite{aps} in order to equip the description with a proper infrared regulator removal limit and thus give robustness to the physical predictions of these symmetry reduced models. On the level of midisuperspace models it has been already considered in spherically symmetric gravity in Ref. \cite{chiouBH}, although there the studies have been restricted to the heuristic level of the so-called \emph{semiclassical effective dynamics} \cite{sv-eff} only and no attempt to build a quantum description was made.

The key idea of the scheme is based on a set of properties of reduced loop quantization for symmetric spacetimes. These properties can be summarized in: $(i)$ the existence of distinguished geometrical (fiducial) structures provided by the symmetries that allow a partial diffeomorphism gauge fixing (in consequence providing a physical meaning to the embedding data)
\footnote{In the standard diffeomorphism invariant formulation of LQG the embedding data is averaged out and has no physical meaning.}, 
$(ii)$ nonexistence of well defined curvature operators which thus have to be built via \emph{Thiemann's regularization} process \cite{t-qsd5}. The technical implementation of $(ii)$ in symmetry reduced treatment involves approximating the classical curvature by holonomies of finite closed loops along the edges generated by the Killing vectors, whereas $(i)$ fixes the embedding data as determined by the physical area of the chosen regularization loop. Following \cite{aps} that area is heuristically fixed as the $1$st nonzero eigenvalue of the full LQG area operator\footnote{Outside of isotropic models 	this procedure is substantially more involved. See the discussion in \cite{awe-b1}.}, known as the \emph{area gap}  and denoted by $(\ell_{\rm Pl}\lambda)^2$, with $\ell_{\rm Pl}^2=\hbar G$ the (square of the) Planck length.

Here, the regularization loops, built at each vertex $v_j$ of the graph, are squares. The length (with respect to certain fiducial metric) $\rho_j$ of each side is fixed such that the physical area of the loop equals exactly $(\ell_{\rm Pl}\lambda)^2$.
Following LQG, these areas are determined by the action of the area operator $\hat{A}$---with eigenvalues $\ell_{\rm Pl}^2 a_j(\vec{k})$---built in turn out of the flux operators (see Ref. \cite{qsd} for details). The area of the surface enclosed by the loop (at vertex $j$) with fiducial sides $\rho_j$ is given by  $\ell^{2}_{\rm Pl}\rho^{2}_{j} a_j(\vec{k})$.\footnote{Here we follow the convention where $\lambda$ and $a_j$ are dimensionless.}
In this situation, the action of the point holonomies of length $\rho_{j}$ will produce a shift in a state $|\mu_j\rangle$ by a ``length'' which depends on the phase space variables. Therefore, it will be convenient to adopt a more appropriate state labeling $\mu_{j} \rightarrow \nu_{j}=\sqrt{a_j(\vec{k})}\mu_{j}/\lambda$. 

After this relabeling, the kinematical Hilbert space is constructed as the closure of the space spanned by Eq. \eqref{eq:Hkin-base} with respect to the inner product $\langle\vec{k},\vec{\nu}|\vec{k}',\vec{\nu}'\rangle=\delta_{\vec{k}\vec{k}'}\delta_{\vec{\nu}\vec{\nu}'}$, further generalized by the rule that basis states belonging to different graphs are mutually orthogonal\footnote{This is actually guaranteed if we allow vanishing labels for the quantum numbers $\vec{k}$ and $\vec{\nu}$.}. 

Once the kinematical Hilbert space is constructed we promote a set of classical variables to operators in which we follow (modulo minor refinements) the proposals in \cite{b-spher,bs-spher}. These are:
\begin{enumerate}
	\item The ``triad component'' operator, in the precise definition following the ideas of LQC \cite{abl-lqc} 
		as the (appropriately rescaled) flux of the
		triad component ${\cal E}$ across the Killing orbit surface, which can be done due to the symmetries of the model.
		Its action takes a very simple form
		\begin{equation}\label{eq:Edef}
			{\hat{\cal E}(\theta) } |\vec{k},\vec{\nu}\rangle
			= 2\gamma\ell_{\rm Pl}^2 \pi^{-1} k_{j(\theta)} |\vec{k},\vec{\nu}\rangle,
		\end{equation}
		where $j(\theta)$ is the index corresponding to the edge $e_j$ going through $\theta$.
		We extend the definition of $\hat{\cal E}(\theta)$ to the 
		vertices of the graph by considering the contributions of both edges connecting the vertex (with weight $1/2$), i.e. $k_{j(\theta_j)}=(k_{j}+k_{j-1})/2$.
		It is also convenient to define an operator 
		\begin{equation}\label{eq:Ejdef}
			{\hat{\cal E}}_j |\vec{k},\vec{\nu}\rangle
			= 2\gamma\ell_{\rm Pl}^2 \pi^{-1} k_{j} |\vec{k},\vec{\nu}\rangle ,
		\end{equation}
		corresponding to the flux over a surface intersected by the edge $e_j$, and 
		\begin{equation}\label{eq:Ejdef-v}
			{\hat{\cal E}^v}_j |\vec{k},\vec{\nu}\rangle
			= \gamma\ell_{\rm Pl}^2 \pi^{-1} (k_{j}+k_{j-1}) |\vec{k},\vec{\nu}\rangle ,
		\end{equation}
		if the intersection is at a vertex.
		
		At this moment it is necessary to recall that associating the operator to a particular edge/vertex, although it is standard in midisuperspace models 
		and spin foam approaches, it does not follow from the standard quantization procedure as one cannot construct a classical observable distinguishing 
		it. Instead, it constitutes an additional nontrivial component of the implemented treatment.
	\item The operator corresponding to the area of the Killing orbit surface, defined as
		\begin{equation}
			\hat{A}(\theta) |\vec{k},\vec{\nu}\rangle = \sqrt{{\cal E}^2(\theta)} |\vec{k},\vec{\nu}\rangle .
		\end{equation}
		For $\theta$ lying in the interior of an edge $e_j$ its action reads
		\begin{equation} 
			\hat{A}(\theta) |\vec{k},\vec{\nu}\rangle = 2\gamma\ell_{\rm Pl}^2 \pi^{-1} |k_{j(\theta)}| |\vec{k},\vec{\nu}\rangle , 
		\end{equation}
		whereas on the vertex $v_j$ it has (as in full LQG) contributions from both edges
		\begin{equation} \label{eq:area-op}
			\hat{A}_j |\vec{k},\vec{\nu}\rangle 
			= \ell_{\rm Pl}^2 a_j(\vec{k}) |\vec{k},\vec{\nu}\rangle 
			= \gamma\ell_{\rm Pl}^2 \pi^{-1} ( |k_j| + |k_{j-1}| ) |\vec{k},\vec{\nu}\rangle .
		\end{equation}
	\item The operator $\hat{V}({\cal I}) 
		= \int_{{\cal I}}\rd\theta \widehat{\sqrt{A}}(\theta)\hat{E}^{x}(\theta)$ corresponding to the
		``volume of a region'' ${\cal I} \subset {\cal B}$, of which action reads
		\begin{equation}\label{eq:Vdef}
			\hat{V}({\cal I}) |\vec{k},\vec{\nu}\rangle =  2\gamma \lambda \ell_{\rm Pl}^3 \pi^{-1}
				\sum_{v_j\in {\cal I}} |\nu_{j}| |\vec{k},\vec{\nu}\rangle.
		\end{equation}
		Again, it is convenient to introduce the volume operator 
		associated to a vertex $v_j$ by choosing the interval ${\cal I}_j$ so that $v_j$ is the only vertex it contains. 
		Its action reads
		\begin{equation}\label{eq:Vjdef}
			\hat{V}_j |\vec{k},\vec{\nu}\rangle = 2\gamma \lambda \ell_{\rm Pl}^3 \pi^{-1} |\nu_j||\vec{k},\vec{\nu}\rangle .
		\end{equation}
	\item The point holonomy operator  \mbox{$\hat{{\cal N}}_{j}:=\widehat{\exp}(i\rho_j K_x(\theta_j))$} defined on a vertex $v_j$ has an action
		on the corresponding subspace given by
		\begin{equation}\label{eq:Ndef}
			\hat{{\cal N}}_{\rho_j}|\nu_j\rangle = |\nu_j+1\rangle
		\end{equation}
		being the identity on the remaining subspaces.
\end{enumerate}
For mathematical convenience, adopting the construction commonly used when dealing with the volume operator in LQG \cite{al-vol} and in the context of 
midisuperspace models postulated originally in Eq. (29) of \cite{bs-ham} we can think of the operator \eqref{eq:Vdef} as the integral of the distribution valued ``volume form operator density''
\begin{equation}\label{eq:V-dist}
	\hat{V}(\theta) |\vec{k},\vec{\nu}\rangle
	= 2\gamma\lambda \ell_{\rm Pl}^3 \pi^{-1} \sum_{\nu_j\in g} \delta(\theta-\theta_j)\nu_{j(\theta)} 
	|\vec{k},\vec{\nu}\rangle .
\end{equation}
Note however that, unlike Eq. \eqref{eq:Vdef}, this ``volume density'' is not in itself a well defined operator.
In the remaining part of the paper we will omit its $\theta$-dependence unless otherwise specified.

Due to the absence (thanks to the abelianization procedure of Sec.~\ref{sec:abelian}) of the variable ${\cal A}$ in the Hamiltonian constraint, it will be not necessary to construct the operator corresponding to the holonomy along edges $e_j$.

The Hilbert space and basic operators defined in this section will serve as an arena for the second step in the Dirac program, promoting the (relevant) classical constraint to an operator and finding its kernel. Following the treatment of full LQG this procedure will be applied to the Hamiltonian constraint.

\section{The scalar constraint}\label{sec:qham} 

As in all applications of the loop quantization to either full GR or symmetry reduced models, the components of the Hamiltonian constraint cannot be directly promoted to operators as most of them do not exist on the kinematical level of quantization. It is necessary to first express or approximate them via variables of which quantum counterparts we have at our disposal (which here means the operators given in the Eqs. \eqref{eq:Edef} and \eqref{eq:Vdef}, and their powers) or their Poisson brackets. This method is known in the literature as the Thiemann's regularization \cite{qsd,t-qsd5}. 
Subsequently, the quantum constraint operator will be defined by directly promoting the components of the regularized constraint to operators.

As in other loop quantized models, the procedure mentioned above permits a series of ambiguities related with either the details of the regularization (which is not uniquely defined) or the factor ordering. These ambiguities are fixed in the following way:
\begin{itemize}
	\item To express the connection $K_x$ in terms of holonomies we employ the simplest possible approximation: 
		by the difference of a holonomy and its inverse. Mathematically this procedure amounts to the substitution 
		$K_x\to\sin\left(\rho_{j} K_x\right)/\rho_{j}$.
	\item The inverse volume is regularized via the application of Thiemann's scheme proposed by Bojowald \cite{b-vol}
		and next adapted to the improved dynamics scheme in Ref. \cite{aps}.
	\item The chosen factor ordering is a straightforward generalization of the so-called \emph{MMO} scheme 
		\cite{mmp,mmo-frw,mmp-presc}, which ensures decoupling distinct orientations of $\nu_j$ and the ``classical singularity states''---basis
		states $|\vec{k},\vec{\nu}\rangle$ for which any component of $\vec{k}$ or $\vec{\nu}$ equals zero---from
		the dynamics. 
	\item Following the invariance of the Hamiltonian constraint operator with respect to the orientation-reflection symmetry, 
		the operator $\sqrt{{\cal E}}$ is represented as $[{\cal E}^2]^{1/4}$, thus becoming a power of an area operator.
	\item The spatial derivative $\partial_{\theta}{\cal E}$ has been promoted to an operator after the following simple observation: $(i)$ the operator ${\cal E}(\theta)$ is diagonal 
		\begin{equation}\
		  \widehat{\partial_{\theta}{\cal E}} |\vec{k},\vec{\nu}\rangle = \partial_{\theta}{\cal E} |\vec{k},\vec{\nu}\rangle
		\end{equation}
		with each nontrivial coefficient ${\cal E}(\theta)$ constant on each edge of the graph and discontinuous jumps on 
		the graph vertices. Thus, calculating directly the derivative of this coefficient on the embedded graph 
		would yield the following result
		\begin{equation}\label{eq:dEdef}
			\partial_{\theta}{\cal E} = 2\gamma\ell_{\rm Pl}^2 \pi^{-1} \sum_{v_j} \Delta k_j \delta(\theta-\theta_j) , 
		\end{equation}
		where $\Delta k_j := (k_j - k_{j-1})$, $\theta_j$ is the position of the vertex $v_j$ and we follow 
		the numbering convention where $v_j$ is the 
		left-hand side boundary of the edge $e_j$. Since it is a distribution, taking it as a coefficient of a 
		diagonal operator would not give a well defined result. It does however provide a correct definition of a smeared 
		operator (similarly to $V(\mathcal{I})$)
		\begin{equation}
			\begin{split}\label{eq:dEIdef}
				\widehat{\partial_{\theta}{\cal E}}(\mathcal{I}) |\vec{k},\vec{\nu}\rangle
				:= 2\gamma\ell_{\rm Pl}^2 \pi^{-1} \left[ \int_{\mathcal{I}} \rd\theta \partial_{\theta}{\cal E} \right] |\vec{k},\vec{\nu}\rangle  \\
				= 2\gamma\ell_{\rm Pl}^2 \pi^{-1} \left[ \sum_{v_i\in\mathcal{I}} \Delta k_j \right] |\vec{k},\vec{\nu}\rangle . 
			\end{split}
		\end{equation}
		This, in turn, by selecting $\mathcal{I} = \mathcal{I}_j$, such that it contains only the vertex $v_j$,  
		allows us to define the ``vertex difference operator''
		\begin{equation}\label{eq:dEjdef}
		  \widehat{[\partial_\theta \cal E]}_j |\vec{k},\vec{\nu}\rangle 
			= 2\gamma\ell_{\rm Pl}^2 \pi^{-1}\Delta k_{j} |\vec{k},\vec{\nu}\rangle .
		\end{equation}
		A quite surprising (and counterintuitive for the quantum counterpart of the spatial derivative operator) 
		property of such definition is its independence of the action on the coordinate length of the involved 
		edges of the graph.
	\item Since the derivatives of operators defined on the open intervals on ${\cal B}$ cannot be defined in a straightforward 
		way, thus the ``global'' spatial derivative present in the Abelian Hamiltonian constraint (see Eq.~\eqref{eq:H-tot-class}) 
		has to be regularized. One of the ways (not necessarily optimal)
		to achieve that is to apply the mathematical shortcut discussed in the context of the volume operator $V({\cal I})$ when 
		the ``distributional operator'' $V(\theta)$ has been defined as in Eq. \eqref{eq:V-dist}. Such procedure will replace the derivative in question 
		with the difference of the ``interior'' operators between two vertices of a given edge.
\end{itemize}
The application of the listed choices results in the operator form of the Hamiltonian constraint taking the form
\begin{equation}\label{eq:quant-scalar-constrh0}
	\hat{\tilde{H}}(N)=\ell_{\rm Pl}\sum_{j} N_j\hat P 
		\left[a_{j-1}^{3/2}\hat{ h}_{j-1}-a_j^{3/2}\hat{h}_{j}\right]\hat P,
\end{equation}
where 
\begin{subequations}\label{eq:qH-aux}\begin{align}
  &\hat P|\vec{k},\vec{\nu}\rangle = \prod_{j}\left[{\rm sgn}(k_j){\rm sgn}(\nu_j)\right]^2|\vec{k},\vec{\nu}\rangle , \\                    
	&\hat h_j = \widehat{\left[\frac{1}{V}\right]}_j^{1/2} \!\! \left( 2 \hat\Omega^2_j
		-\frac{2\gamma^2}{\pi^{2}} \widehat{\left[\frac{1}{V}\right]}_j \ell_{\rm Pl}^2(\ell_{\rm Pl}^2\Delta k_j)^2\right)\!\!\widehat{\left[\frac{1}{V}\right]}_j^{1/2} , \label{eq:hj-def} \\
	\begin{split}
		&\widehat{\left[\frac1{V}\right]}_j^{1/2} |\vec{k},\vec{\nu}\rangle
		= \frac{b(\nu_j)}{(2\gamma \lambda \ell_{\rm Pl}^3\pi^{-1})^{1/2}} |\vec{k},\vec{\nu}\rangle , \\
		&\hphantom{=} \qquad\  b(\nu_j):=||\nu_j+1|^{1/2}-|\nu_j-1|^{1/2}| ,
	\end{split} \\
	\begin{split}	 
		&\hat{\Omega}_j= \frac{1}{{4i\gamma\lambda}}|\hat{V}_j|^{1/4}\big[\widehat{{\rm sgn}(\nu_j)}\big(\hat {\cal N}_{\rho_{j}}^2-\hat {\cal N}_{\rho_{j}}^{-2}\big) \\
		&\hphantom{=}\ 
		+\big(\hat {\cal N}_{\rho_{j}}^2-\hat {\cal N}_{\rho_{j}}^{-2}\big)\widehat{{\rm sgn}(\nu_j)}\big]|\hat{V}_j|^{1/4} .
	\end{split}
\end{align}
\end{subequations}

An analogous Hamiltonian constraint operator, taking very similar form, has been already studied in spherically symmetric spacetimes \cite{BH-2}. It commutes with itself and is free of anomalies. Besides, due to the factor ordering choice, the states with either $k_j=0$ or $\nu_j=0$, or both, are trivially annihilated by the constraint. In consequence, they will be irrelevant for the dynamics, and can be safely removed from the space of solutions to the constraint. As a consequence, the quantum theory is then able to cure those coordinate and curvature singularities
\footnote{It is well established that the considered Gowdy model possesses a curvature singularity when the area of the Killing orbits vanishes \cite{Berger:2002st}, i.e., when $\mathcal{E}$ vanishes.} 
arising either at ${\cal E}=0$ and/or $E^x=0$.

The quantum Hamiltonian constraint constructed here will be next used (after suitably dealing with the diffeomorphism group)
to construct the physical Hilbert space, further allowing to study the dynamical sector of the theory.

\section{Physical Hilbert space, dynamical sector}
\label{sec:phys}

At this point we have at our disposal the kinematical Hilbert space, i.e., well-defined operator(s) acting on that space and the operator form of the Hamiltonian constraint. The remaining tasks in completing the quantization program are: finding the physical Hilbert space as the (dual of the) kernel of the constraint operators and constructing a set of physically meaningful observables acting on this space. Its first step: solving the constraints requires an additional effort as at present we do not have (nor we intend to build) a quantum diffeomorphism constraint operator. While recently a construction of such quantum diffeomorphism generator has been proposed in the full theory \cite{mad}, the standard treatment of diffeomorphisms in loop quantization just uses the finite diffeomorphisms. Here we employ the same philosophy, applying the constraint-solving program of the full LQG. There, the constraints are implemented in hierarchy: $(i)$ first, the diffeomorphism-invariant Hilbert space is constructed out of the kinematical one by the so-called \emph{group averaging procedure} \cite{ga}, $(ii)$ next the actual physical Hilbert space is defined as a space annihilated by the Hamiltonian constraint operator acting on the diffeomorphism-invariant space
\footnote{A condition necessary for this step is that the Hamiltonian operator is well defined on this space, which is exactly the case in the full LQG.}. 
We repeat this exact procedure in the context of our model. 

First, let us identify the diffeomorphism-invariant sector of the theory via group averaging.

\subsection{Averaging over the spatial diffeomorphisms}\label{sec:gave-diff}

Consider a general situation of a compact group of transformations (classically generated by a set of constraints) represented in the quantum theory by unitary operators acting on certain Hilbert space. In such situation one is usually interested in finding a sector of the quantum theory under study, which is invariant with respect to these transformations. It usually lies in the algebraic dual of a dense subspace of the original Hilbert space and roughly speaking consists of ``averages'' of states transformed over the whole transformation group. Because of this principal idea the procedure of building such states is known as the \emph{group averaging technique} \cite{ga,m-gave1,*m-gave2,*m-gave3,*m-gave4}. In a precise sense the averaging is achieved by constructing for the group $K$ of unitary transformations $\hat{U}(\alpha):\Hil\to\Hil$ parametrized by the index $\alpha$ (such that $\rd\alpha$ is the bi-invariant 
measure on $K$) the antilinear rigging map $\eta:\Hil\to\Hil^\star$ such that
\begin{equation}\label{eq:ga-s}
	\Hil^\star\ni\eta|\psi\rangle = \left(\int_{K} \rd\alpha \hat{U}(\alpha)|\psi\rangle\right)^{\dagger} 
	= \int_K \rd \alpha \langle\psi|\hat{U}^{-1}(\alpha) . 
\end{equation}
The states $|\phi) = \eta|\psi\rangle$ (when nontrivial) span a transformation-invariant Hilbert space with induced inner product
\begin{equation}\label{eq:ga-ip}
	(\eta\psi|\eta\chi) = \int_K \rd \alpha \langle\chi|\hat{U}^{-1}(\alpha)|\psi\rangle . 
\end{equation}

Our goal in this section is to define such rigging map averaging over the group of spatial diffeomorphisms of the Cauchy slice of the LRS Gowdy spacetime. This group is relatively large, however, as we will see below, for the purpose of averaging it can be reduced to a small (in fact finite dimensional) subgroup. To perform this reduction we first note that, due to a natural requirement, that the transformations preserve the symmetries of the selected class of spacetimes, the group of diffeomorphisms reduces to a subgroup of diffeomorphisms on the reduced manifold ${\cal B}$. We will denote this subgroup by ${\rm Diff}_{{\cal B}}$.

Each diffeomorphism $\varphi$ acts on a given graph $|g,\vec k,\vec \nu \rangle$ yielding a new state $|g_\varphi,\vec k_\varphi,\vec \nu_\varphi \rangle$ such that it drags the vertices of the graphs (together with all the points of ${\cal B}$) in such a way that the sets $\vec k_\varphi=\vec k$ and $\vec \nu_\varphi=\vec \nu$. Then it is the position of any point in ${\cal B}$ that is dragged to the new position $\tilde \theta$, while preserving the order of the points, that is: $\theta>\theta' \Rightarrow \tilde \theta>\tilde \theta'$. Thus $\varphi$ induces a unitary operator $\hat U_\varphi$ on the kinematical Hilbert space such that $|g_\varphi,\vec k_\varphi,\vec \mu_\varphi \rangle=\hat U_\varphi|g,\vec k,\vec \mu \rangle$. We will parametrize the group of the operators with the index $\beta$, such that $\rd\beta$ provides the bi-invariant measure on the group.

Next, we note that due to the orthogonality (with respect to the kinematical inner product) of the disjoint graphs, for the evaluation of the integrals in Eqs. \eqref{eq:ga-s} and \eqref{eq:ga-ip} it is sufficient to consider an action of the diffeomorphism transformation group on each single closed graph separately.
Consider now a subgroup ${\rm Diff}_{{\cal B}g}$ of diffeomorphisms preserving the positions of all the edges and vertices of a particular graph. The action of the unitary transformation operator $\hat{U}(\beta)$ corresponding to each element of this subgroup is the identity operator on $\Hil_{{\rm kin}}$. Furthermore, we can group all the elements $|g,\vec{k},\vec{\nu}\rangle$ of the basis of $\Hil_{{\rm kin}}$ into classes of equivalence with respect to transformations from ${\rm Diff}_{{\cal B}g}$. These classes of equivalence will span a Hilbert space $\Hil_{\rm aux}$ with an inner product induced in a straightforward way from $\Hil_{\rm kin}$ since the latter is invariant with respect to parametrizations of graph edges by $\theta$. The only difference of the new space with respect to $\Hil_{{\rm kin}}$ is that now the information about the parametrization of the graph edges (which in embedded graph are curves on ${\cal B}$) is removed.
As a consequence we can reformulate the integrals in Eqs. \eqref{eq:ga-s} and \eqref{eq:ga-ip} as integrals over the 
quotient group ${\rm Diff}_{{\cal B}}/{\rm Diff}_{{\cal B}g}$ of the transformations acting on $\Hil_{{\rm aux}}$,
\begin{subequations}\label{eq:ga}\begin{align}
  \eta|\psi\rangle &= \int_{{\rm Diff}_{{\cal B}}/{\rm Diff}_{{\cal B}g}} \rd \beta \langle\psi|\hat{U}^{-1}(\beta) , \\
	(\eta\psi|\eta\chi) &= \int_{{\rm Diff}_{{\cal B}}/{\rm Diff}_{{\cal B}g}} 
		\rd \beta \langle\chi|\hat{U}^{-1}(\beta)|\psi\rangle_{{\rm aux}} .
\end{align}\end{subequations}

The considered quotient group is already finite-dimensional. The transformations are just shifts of positions of the edges of the graph preserving their order. To characterize the group we note that, once the information about the parametrization of edges is removed (which happened in going to the auxiliary Hilbert space) each embedded graph can be characterized (up to a global phase rotation) by a sequence of $N$ fiducial lengths $\Delta\theta_j$ of the graph edges, thus be represented as a point on the surface $\sum_j \Delta\theta_j = 2\pi$ in the space $(\mathbb{R}^{+})^N$. Then (after further dividing by the $U(1)$ subgroup of the above-mentioned rigid rotations) each transformation from the considered quotient group is just a shift on this surface.

To build a meaningful rigging map we select a discrete measure $\rd\beta$. Then the integrals \eqref{eq:ga} become uncountable sums and can be calculated quite easily. Indeed, the inner product in the diffeomorphism invariant Hilbert space becomes now 
\begin{equation}
	(\eta\psi|\eta\chi)_{{\rm diff}} 
	= \sum_{\varphi \in {\rm Diff}_{{\cal B}}/{\rm Diff}_{{\cal B}g}} \langle \psi| \hat{U}^{\varphi}\chi \rangle .
\end{equation}
Since each transformation preserves the labels $\vec{k}, \vec{\nu}$, order of vertices and edges (and orientation of the latter) of the auxiliary state, the basis of $\Hil_{{\rm diff}}$ can be constructed out of states $|\vec{k}, \vec{\nu}\rangle$ now understood as living on the abstract graph (with only ordering of the vertices instead of the embedding). The diffeomorphism invariant inner product will mathematically take the same form (modulo vertex positions) as the kinematical one 
\begin{equation}
	\langle \vec{k},\vec{\nu}|\vec{k}',\vec{\nu}'\rangle = \prod_{j=1}^N \delta_{k_j,k'_j}\delta_{\nu_j,\nu'_j} . 
\end{equation}

Once $\Hil_{\rm diff}$ has been constructed, the next step is to build analogs of the elementary operators (originally acting in $\Hil_{{\rm kin}}$) which will be well defined on a dense domain in $\Hil_{{\rm diff}}$. For that we again employ the group averaging, defining for a given $\hat{O}:\mathcal{D}\subset\Hil_{{\rm kin}} \to \Hil_{{\rm kin}}$ an operator $\hat{O}_{{\rm diff}}:\mathcal{D'}\subset\Hil_{{\rm diff}} \to \Hil_{{\rm diff}}$ (where $\mathcal{D},\mathcal{D'}$ are dense domains in their respective Hilbert spaces) in the following way
\begin{equation}\label{eq:ga-op}
	\hat{O}_{{\rm diff}} := \int_{{\rm Diff}_{{\cal B}}/{\rm Diff}_{{\cal B}g}} \rd\beta \hat{U}(\beta)^{-1}\hat{O}\hat{U}(\beta) . 
\end{equation}
An application of this procedure to Eqs. \eqref{eq:Ndef} and \eqref{eq:dEjdef} yields operators which are mathematically identical to their kinematical predecessors, although now they act on the labels of the 
\emph{abstract} graphs. However, the families of operators  defined by Eqs. \eqref{eq:Edef}, \eqref{eq:Vdef} and \eqref{eq:dEdef} will not provide useful definitions as they are parametrized by either points or regions on ${\cal B}$. Instead, one needs to take their versions in Eqs. \eqref{eq:Ejdef}, \eqref{eq:Vjdef} and \eqref{eq:dEjdef}, parametrized by the vertex label $j$ which plays the role of the ordering index of the graph elements.
Thus, we end up with the following set of diffeomorphism invariant elementary operators
\begin{subequations}\label{eq:op-diff}\begin{align}
	\hat{{\cal E}}_j |\vec{k},\vec{\nu}\rangle &= 2\gamma \ell_{\rm Pl}^2\pi^{-1} k_j |\vec{k},\vec{\nu}\rangle , \\
	\hat{A}_j |\vec{k},\vec{\nu}\rangle &= \gamma \ell_{\rm Pl}^2\pi^{-1}(|k_j|+|k_{j-1}|) |\vec{k},\vec{\nu}\rangle , \label{eq:A-def} \\
	\hat{V}_j |\vec{k},\vec{\nu}\rangle &= 2\gamma \lambda \ell_{\rm Pl}^3\pi^{-1}|\nu_j| |\vec{k},\vec{\nu}\rangle , \\
	\widehat{\partial_{\theta}{\cal E}}_j |\vec{k},\vec{\nu}\rangle &= 2\gamma \ell_{\rm Pl}^2\pi^{-1} (k_j-k_{j-1}) |\vec{k},\vec{\nu}\rangle 
\end{align}\end{subequations} 
and 
\begin{equation}\label{eq:N-diff}
	\hat{{\cal N}}_{\rho_j} |\vec{k},(\nu_1,\ldots,\nu_j,\ldots,\nu_N)\rangle = |\vec{k},(\nu_1,\ldots,\nu_j+1,\ldots,\nu_N)\rangle . 
\end{equation}
Given these operators, one can immediately write a version of the quantum Hamiltonian constraint defined on the dense domain of $\Hil_{{\rm diff}}$ by averaging [via Eq. \eqref{eq:ga-op}] the operators in Eqs. \eqref{eq:quant-scalar-constrh0} and \eqref{eq:qH-aux}. This procedure will yield a constraint in which all the elementary operators are replaced with their diffeomorphism invariant counterparts defined in Eqs. \eqref{eq:op-diff} and \eqref{eq:N-diff}.
This result is critical in performing the next step in building the physical Hilbert space---finding the kernel of this constraint (as the only one remaining).

\subsection{Solutions to the Hamiltonian constraint}

Unlike the diffeomorphisms, in loop approaches the Hamiltonian constraint is implemented directly as a generator, thus (as stated earlier) solving it corresponds to finding the states annihilated by its quantum counterpart. Technically, it amounts to solving the equation for the (generalized) wave function $\psi$ in the algebraic dual of a suitable dense subset of $\Hil_{\diff}$ 
\begin{equation}\label{eq:ker}
	\begin{split}
		\forall |\chi\rangle &\in\Hil_{{\rm diff}}:\ (\Psi|\hat{\tilde{H}}|\chi\rangle = 0 , \\ 
		(\Psi| &=\sum_{\vec k}\sum_{\vec \nu}\langle \vec{k},\vec{\nu}|\psi(\vec k,\vec{\nu}) .
	\end{split}
\end{equation}
This can be again achieved by the group averaging technique. Here, however, unlike in the case of differomphisms we have at our disposal the generator of the transformations (in this case time reparametrizations). Thus, the rigging map can be written as 
\begin{equation}\label{eq:H-ave}
	\eta_H|\Psi\rangle = \int_{\re^{N}} \rd^N \vec{N} \left( e^{i\hat{\tilde{H}}(\vec{N})} 
		|\Psi\rangle_{{\rm diff}} \right)^{\dagger} .
\end{equation}
In order to construct this map explicitly we will perform the spectral decomposition of $\hat{\tilde{H}}$. First, we note that this operator [see \eqref{eq:quant-scalar-constrh0}] is a linear combination of mutually commuting component operators $\hat{h}_j$ [defined in \eqref{eq:hj-def}], which thus can be simultaneously diagonalized. It is thus enough to perform the spectral decomposition of each $\hat{h}_j$. To do so we introduce a set of auxiliary operators $\hat{h}_{m,n} : \BohrOne \to \BohrOne$ (where $m,n\in\mathbb{Z}$) of the mathematical form analogous to 
\eqref{eq:hj-def}
 \begin{subequations}\label{eq:hred}
\begin{align}
	&\tilde h_{m,n} = \widehat{\left[\frac{1}{V}\right]}^{1/2} \!\! \left( 2 \hat\Omega^2
		-\frac{2\gamma^2}{\pi^{2}} \widehat{\left[\frac{1}{V}\right]} \ell_{\rm Pl}^2(\ell_{\rm Pl}^2(m-n))^2\right)\!\!\widehat{\left[\frac{1}{V}\right]}^{1/2} , 
\\
 	\begin{split}
 		&\widehat{\left[\frac1{V}\right]}^{1/2} |\nu\rangle
 		= \frac{b(\nu)}{(2\gamma \lambda \ell_{\rm Pl}^3\pi^{-1})^{1/2}} |\nu\rangle , \\
 		&\hphantom{=} \qquad\  b(\nu):=||\nu+1|^{1/2}-|\nu-1|^{1/2}| ,
 	\end{split} \\
 	\begin{split}	 
 		&\hat{\Omega} = \frac{1}{{4i\gamma\lambda}}|\hat{V}|^{1/4}\big[\widehat{{\rm sgn}(V)}\big(\hat {\cal N}_{\rho}^2-\hat {\cal N}_{\rho}^{-2}\big) \\
 		&\hphantom{=}\ 
 		+ (\hat {\cal N}_{\rho}^2-\hat {\cal N}_{\rho}^{-2}\big)\widehat{{\rm sgn}(V)}\big]|\hat{V}|^{1/4} ,
 	\end{split}
\end{align}
\end{subequations}
where $\hat{\cal N}_{\rho}|\nu\rangle=|\nu+1\rangle$. Since in the eigenvalue problem $\hat{h}_j|\psi_j\rangle = \omega_j|\psi_j\rangle$ the dependence on $\vec{k}$ is algebraic only one can split it onto a set of independent eigenvalue problems on $\BohrOne$ involving the auxiliary 
operators $\hat{h}_{m,n}$ and parametrized by values of $\vec{k}$. Precisely, we realize this split via introducing the auxiliary Hilbert space $\Hil_{\nu}:=\BohrOne^N$ (corresponding to the vertex degrees of freedom) and a set of the projection/embedding operators 
\begin{subequations}\begin{align}
	\hat{P}_{\vec{k}'}&: \Hil_{\diff} \to \Hil_{\diff} , & 
	\hat{P}_{\vec{k}'} |\vec{k},\vec{\nu}\rangle &= \prod_j \delta_{k_j,k'_j} |\vec{k},\vec{\nu}\rangle , \\
	\hat{R}&:\Hil_{\diff} \to \Hil_{\nu} , &
	\hat{R}|\vec{k},\vec{\nu}\rangle &= |\vec{\nu}\rangle , \\
	\hat{Q}_{\vec{k}}&: \Hil_{\nu} \to \Hil_{\diff} , &
	\hat{Q}_{\vec{k}}|\vec{\nu}\rangle &= |\vec{k},\vec{\nu}\rangle .
\end{align}\end{subequations}
The component operators $\hat{h}_j$ can now be written as
\begin{equation}
	\hat{h}_j = \hspace{-3mm} \sum_{\vec{k}'\in (\mathbb{Z}^\star)^N} \hspace{-3mm} 
		\hat{Q}_{\vec{k'}} \hspace{-1mm} \left[\left(\prod_{j'=1}^{j-1} \id \right)\! \times \hat{h}_{k'_j,k'_{j-1}} \times \! \left(\prod_{j'=j+1}^{N} \id\right) \right] 
		\hat{R} \hat{P}_{\vec{k}'} ,
\end{equation}
where $\mathbb{Z}^\star:= \mathbb{Z}\setminus\{0\}$ and the product in the parenthesis must be understood as a Cartesian product. The eigenvalue problem reduces to the set of equations for the eigenfunctions $\chi_{\omega,k_j,k_{j-1}}$ 
\begin{equation}\label{eq:aux-eip}
	[\hat{h}_{k_j,k_{j-1}} \chi_{\omega,k_j,k_{j-1}}](\nu) = \omega_{k_j,k_{j-1}}\chi_{\omega,k_j,k_{j-1}}(\nu) .
\end{equation}

The properties of the operators $\hat{h}_{m,n}$ have been analyzed in detail in Appendix~\ref{app:disp-rel}. They are essentially selfadjoint (barring an extreme fine-tuning of the Barbero--Immirzi parameter), the spectrum of each of them is nondegenerate and its continuous part is ${\rm Sp}_{{\rm cont}}(\hat{h}_{m,n}) = [0, 2(\gamma\lambda)^{-2}] =: I_{\omega}$.  

Consider now a set of Dirac delta normalized solutions to \eqref{eq:aux-eip}, denoted further by $\tilde{e}_{\omega,m,n}(\nu)$. Then, the functions of the form
\begin{equation}\label{eq:e-def-2}
	e_{\omega,j,m,n}(\vec{k},\vec{\nu}) = \delta_{m,k_j}\delta_{n,k_{j-1}} f(\vec{v}_{(j)}) \tilde{e}_{\omega,m,n}(\nu_j) ,
\end{equation}
are the (normalized) eigenfunctions of $\hat{h}_j$, where $f(\vec{v}_{(j)})$ is any normalized function on the $(N-1)$-dimensional space of vectors such that $\vec{v}$ corresponds to any vector $\vec\nu$ with the $j$-th coordinate removed. As a consequence, the spectrum of $\hat{h}_j$  also has a continuous part ${\rm Sp}_{{\rm cont}}(\hat{h}_{j}) = I_{\omega}$, although now it has a continuous degeneracy (originating from the freedom in the choice of function $f$ in \eqref{eq:e-def-2}). We note however that this degeneracy becomes spurious when we consider the complete Hamiltonian constraint operator.

The mutual eigenfunctions of $\hat{h}_j$ corresponding to the vector of eigenvalues $\vec{\omega}:=(\omega_1,\ldots,\omega_N)^T\in [0,2(\gamma\lambda)^{-2}]^N$ (parametrized by $\vec{k}$) are the linear combinations of products of the (Dirac delta normalized) solutions to \eqref{eq:aux-eip}
\begin{equation}
	e_{\vec{\omega},\vec{k}}(\vec{k}',\vec{\nu}') 
	= \prod_{j=1}^N \delta_{k_j,k_j'} \tilde{e}_{\omega_j,k_j,k_{j-1}}(\nu'_j) . 
\end{equation}
By construction they are also eigenfunctions of $\tilde{a}_j\id$, thus they diagonalize \emph{all} the component operators of $\hat{H}$. 

Consider now the diffeomorphism invariant state $|\Psi\rangle$ decomposed in the above basis
\begin{equation}
	\langle \vec{k}',\vec{\nu}'|\Psi\rangle 
	= \int_{I_{\omega}^N} \rd^N \vec{\omega} \sum_{\vec{k}\in (\mathbb{Z}^\star)^n}
		\tilde{\Psi}(\vec{\omega},\vec{k}) \langle \vec{k}',\vec{\nu}'|e_{\vec{\omega},\vec{k}}) .
\end{equation}
When acting on it, the rigging map \eqref{eq:H-ave} (well defined as $\tilde{\hat{H}}$ is essentially self-adjoint, which follows from essential self-adjointness of the component operators $\tilde{h}_{m,n}$) produces the following physical state
\begin{widetext}
\begin{equation}
	\begin{split}
		\eta_H|\Psi\rangle &= \int \rd^N \vec{t} \int_{I_{\omega}^N} \rd^N \vec{\omega} \sum_{\vec{k}\in (\mathbb{Z}^\star)^n} 
			\exp\left[ i \left(\sum_{j=1}^N t_j 
				(a_{j-1}^{3/2}\omega_{j-1,\vec{k}} - a_j^{3/2}\omega_{j,\vec{k}}) 
			\right) \right] \tilde{\Psi}(\vec{\omega},\vec{k}) (e_{\vec{\omega},\vec{k}}|  \\
		&= \int_{I_{\omega}^N} \rd^N \vec{\omega} \sum_{\vec{k}\in (\mathbb{Z}^\star)^n}
				\prod_{j=1}^N \delta(a_{j-1}^{3/2}\omega_{j-1,\vec{k}} - a_j^{3/2}\omega_{j,\vec{k}}) 
					\tilde{\Psi}(\vec{\omega},\vec{k}) (e_{\vec{\omega},\vec{k}}|.
	\end{split}
\end{equation}
\end{widetext}
As a consequence the group averaging procedure selects out the states of the products $a_{j}^{3/2}\omega_{j,\vec{k}}$ taking the same value on all the vertices of the graph. It is convenient to represent this constraint by introducing the $\vec{k}$-dependent variable $h_{\vec{k}}$ such that
\begin{equation}
	a_{j}^{3/2}\omega_{j,\vec{k}} =: h_{\vec{k}} , 
\end{equation}
which allows to determine the frequencies $\omega_{j,\vec{k}}$ as functions of $\vec{k}$ in the following way
\begin{equation}
	\omega_{j,\vec{k}} = h_{\vec{k}} a_{j}^{-3/2} =: \omega_j(\vec{k},h_{\vec{k}}). 
\end{equation}
Since $\omega_{j,\vec{k}} \in I_{\omega}$ the variables $h_{\vec{k}}$ are non-negative and bounded from above by the function of $\vec{k}$
\begin{equation}
	h_{\vec{k}} \in [0 ,h_\star(\vec{k})] , \quad 
	h_\star(\vec{k}) = \frac{2}{(\gamma\lambda)^2} (\min_j a_j)^{-3/2} , 
\end{equation}
which in turn is bounded by a global constant 
$h_\star(\vec{k}) \leq \frac{\pi^{3/2}}{\sqrt{2}\lambda^2\gamma^{7/2}}$ (due to the decoupling of states with any component of $\vec{k}$ vanishing).

Using the new variables, one can represent the physical states in a slightly simpler form
\begin{equation}\label{eq:phy-fin}
  \langle \vec{k}',\vec{\nu}' | \Psi \rangle 
	= \sum_{\vec{k}\in (\mathbb{Z}^\star)^n} \int_0^{h_\star(\vec{k})} \rd h_{\vec{k}} \tilde{\Psi}(\vec{k},h_{\vec{k}})
	(e_{\vec{\omega}(\vec{k},h_{\vec{k}}),\vec{k}}|\vec{k}',\vec{\nu}'\rangle^\star
\end{equation}
and the physical inner product (induced from $\Hil_{\diff}$ by group averaging) is (unitarily equivalent to)
\begin{equation}\label{eq:scalar-prod}
	\langle \Phi | \Psi \rangle = \sum_{\vec{k}\in (\mathbb{Z}^\star)^n} \int_0^{h_\star(\vec{k})} \rd h_{\vec{k}}
		\tilde{\Phi}^\star(\vec{k},h_{\vec{k}}) \tilde{\Psi}(\vec{k},h_{\vec{k}}) .
\end{equation}

Having the physical Hilbert space at our disposal we can move to the final step of the quantization: defining the physically relevant observables and probing (at least some of) the dynamical properties of the system.

\subsection{Quantum observables}

The Dirac observable capturing the unique global degree of freedom of the system has been defined at the classical level 
in Sec.~\ref{sec:classLRS}, Eq.~\eqref{eq:h1-def}. its functional form (in terms of the classical variables) is identical 
to a component of the Hamiltonian constraint. Thus in order to build its quantum counterpart one can  use the very same methods
applied in sec.~\ref{sec:kin} to quantize that constraint. The result is
\begin{equation}
	\hat{h}_1 = \ell_{\rm Pl} \sum_j \hat{P}\, a_j^{3/2} \hat{h}_j \hat{P} . 
\end{equation}
As $\vec{\omega}(\vec{k},h_{\vec{k}}) \in I_{\omega}^N$ and $\tilde{a}_j$ are explicitly non-negative, this 
observable is also explicitly non-negative on $\Hil_{{\rm phy}}$. 

Its action on the basis elements of $\Hil_{{\rm phy}}$ is very simple---it just multiplies each element by 
an appropriate constant $\ell_{\rm Pl}h_{\vec{k}}$, namely,
\begin{equation}
	\forall |\chi\rangle\in\Hil_{{\rm phy}}\ 
	(e_{\vec{\omega}(\vec{k},h_{\vec{k}}),\vec{k}} | \hat{h}_1 - \ell_{\rm Pl}h_{\vec{k}}\id | \chi \rangle = 0 .  
\end{equation}
This observable is the quantum analogue of the classical phase space function defined in Eq. \eqref{eq:h1-def}. As we mentioned, it does not have an obvious physical interpretation as in the case of spherically symmetric spacetimes (where it encoded the ADM mass). It can be still considered, per analogy, as the ``mass of gravitational shear'', although due to the compactness of the Cauchy slices it looses the connection with any definition of the actual physical mass.

On the other hand, a well defined geometrical meaning can be associated with the operators $\hat{A}_j$ defined in \eqref{eq:A-def} which represent the areas of the 2D Killing surfaces intersecting the vertex $v_j$. Due to the ``ultralocality'' of the Hamiltonian constraint (with respect to its action on $\vec{k}$) they can be easily promoted to operators on $\Hil_{{\rm phy}}$ satisfying the requirements of Dirac observable. Indeed, their action on the basis of $\Hil_{{\rm phy}}$ can be written as
\begin{equation}
	\forall |\chi\rangle\in\Hil_{{\rm phy}}\ 
	(e_{\vec{\omega}(\vec{k},h_{\vec{k}}),\vec{k}} | \hat{A}_j - \ell_{\rm Pl}^2\tilde{a}_j\id | \chi \rangle = 0 .  
\end{equation}
By the same arguments one can define the ``areas'' of the Killing surfaces intersecting the edge $e_j$
\begin{equation}
	\forall |\chi\rangle\in\Hil_{{\rm phy}}\ 
	(e_{\vec{\omega}(\vec{k},h_{\vec{k}}),\vec{k}} | \hat{k}_j - k_j\id | \chi \rangle = 0 .  
\end{equation}
which then are the analogs of the observables defined in the context of spherically symmetric spacetimes in \cite{BH-1,BH-2}. Surprisingly, both sequences of areas are constants of motion.

\subsection{Evolution with $\vec{\nu}$ as time function}

While in this model all the physical information about the state is encoded in the Dirac observables representing constants of motion, to gain a reliable insight into the physics predicted by it one needs to use the set of observables able to reflect the ''changes in time'' of the physical parameters per analogy of changing of the physical properties with respect to cosmic time in the classical theory. For constrained systems it is usually realized by defining the family (or families) of Dirac observables parametrized by a convenient function of phase space variables (serving as the internal clock)---an approach known as the ''evolving constants'' or ''parametrized observables'' \cite{carlo0,carlo1,bianca0,bianca1}. Here, the similarity of the structure of our model with the one describing vacuum Bianchi I spacetimes in LQC allows us to essentially repeat the construction devised in Ref. \cite{mmp-B1-evo}, which consisted of the following steps:
\begin{enumerate}[(i)]
	\item Selection of a suitable phase space variable as the parameter labeling the observables, further called \emph{evolution parameter}, and denoted by $T$.
	\item Defining for each value $T$ of the evolution parameter an ``initial data space'' $\Hil_{T}$, and a unitary\footnote{In 
		this context this means norm preserving.} transformation $\hat{U}_T$ between $\Hil_{\rm phy}$ and $\Hil_{T}$. 
	\item Selecting physically relevant kinematical observables $\{\hat{O}\}$ and casting them as operators acting on suitable domains in $\Hil_T$.
	\item Finally, the evolution is defined via a family of operators $\hat{O}_T := \hat{U}_T^{-1} \hat{O} \hat{U}_T$.
\end{enumerate}
In the case of vacuum Bianchi I the particularly convenient and physically useful choice of internal time was the canonical momentum of the volume. In our model it corresponds to selecting the quantities $\{b_j\}$, canonically conjugate to $\{\nu_j\}$, as the internal time function. 
In this particular model however the asymptotic properties of the eigenfunctions $e_{\vec{\omega},\vec{k}}$ make it much less convenient. Since for large $\nu_j$ the dominant component of each eigenfunction is a combination of ``plane waves'' $e^{i\sigma(k_j,h_j(k_j))\nu}$ the variables $b_j$ freeze, with the value at which they freeze depending on $\vec{k}$. Therefore, we will focus on the choice of $\{\nu_j\}$ as time function. It is worth commenting that the prior analysis of black holes \cite{BH-1,BH-2} upon methods of which our studies are based,
have analyzed the evolution with the connection as time function. Therefore, this work will be the first time that the volume (constructed out of the triad) plays the role of time in the context of Abelianized midisuperspace inhomogeneous models.
	
The choice of $\{\nu_j\}$ as time function shares the difficulties of its counterpart in the homogeneous treatment of the Bianchi I model. Since each eigenfunction $e_{\vec{\omega},\vec{k}}$ is real and furthermore converges to a standing (or, more precisely, reflected) plane wave, building the ``initial data'' Hilbert spaces directly out of the constant slices $\{\nu_j\}$ of $\Psi$ specified in Eq. \eqref{eq:phy-fin} would not lead to a meaningful evolution picture.
\footnote{This can be seen easily on a textbook example of a free particle in $1+1$ dimension with a mirror. There, treating the time as a dynamical variable, taking the Klein-Gordon equation as a constraint and defining the observables $\hat{t}_x$ would yield the family which for a state corresponding to a reflected semiclassical packet in a standard picture  would produce the variance equal to the time separation between each semiclassical packet (right and left moving respectively) and the expectation values being the (constant) average between the ``position'' of each of them} in $t$.
For instance, in Ref. \cite{mmp-B1-evo} it is shown that the evolution between different time slices in this $\nu$-representation can be defined by a suitable unitary map. However, this evolution requires a splitting of the Hilbert space in two sectors and a definition of suitable observables on them. These observables are constructed out of projections of observables in the full physical Hilbert space on each sector and have a clear interpretation (give a good approximation to the original corresponding observables) only in the (low curvature) semiclassical regime.\footnote{Similarly, some of the parametrized observables defined in spherical symmetry have an unambiguous meaning at semiclassical regimes.} Let us see this in more detail.

We start with the following auxiliary states
	\begin{equation}\label{eq:aux-prof-state}\begin{split}
	\Psi_{\vec{\nu}}(\vec{k}) 
	&= \sum_{\vec{k}\in(\mathbb{Z}^\star)^n} \int_0^{h_{\star}(\vec{k})} \rd h_{\vec{k}} \tilde{\Psi}_{\vec{\nu}}(\vec{k},h_{\vec{k}}),
	\end{split}\end{equation}
where
\begin{equation}
\tilde{\Psi}_{\vec{\nu}}(\vec{k},h_{\vec{k}}) = P_{\vec{\nu}}\tilde{\Psi}(\vec{k},h_{\vec{k}})
= e_{\vec{\omega}(\vec{k},h_{\vec{k}})}(\vec{\nu}) \tilde{\Psi}(\vec{k},h_{\vec{k}}) . 
\end{equation}
Since the eigenfunctions $e_{\vec{\omega}(\vec{k},h_{\vec{k}})}(\vec{\nu})$ generically are never vanishing, the transformation $P_{\vec{\nu}}$ is well defined. The new states belong to the Hilbert space $\Hil_{\rm phy}^{\vec{\nu}}$, with an inner product as in $\Hil_{{\rm phy}}$ but with the additional weight $|e_{\vec{\omega}(\vec{k},h_{\vec{k}})}(\vec{\nu})|^{-2}$. Therefore, $P_{\vec{\nu}}$ must be understood as a bijection from the physical Hilbert space to the auxiliary space $\Hil_{\rm phy}^{\vec{\nu}}$. Different choices of $\vec{\nu}$ yields Hilbert spaces $\Hil_{\rm phy}^{\vec{\nu}}$ with different inner products (actually they are time dependent). One can construct bijections between these Hilbert spaces through the operators $Q_{\vec{\nu}_2,\vec{\nu}_1}=P_{\vec{\nu}_2}P^{-1}_{\vec{\nu}_1}:\Hil_{\rm phy}^{\vec{\nu}_1}\to\Hil_{\rm phy}^{\vec{\nu}_2}$. These bijections are ``unitary'' in the sense that they are norm preserving. In this sense, we can provide a notion of evolution. On the other hand, all the spaces $\Hil_{\rm phy}^{\vec{\nu}}$ are actually the same space $\Hil_{\rm phy}$ as the scalar products are equivalent. Therefore one would expect that the identity transformation between spaces at different times is also unitary. However, with our choice of inner products this is not the case. As a consequence the evolution is not unitary in a strict sense. Still, suitable observables can be defined out of the kinematical ones and the transformations $Q_{\vec{\nu}_2,\vec{\nu}_1}$. This notion of evolution is well defined since it keeps all the physical information. Nevertheless, it is not necessary to renounce to a unitary evolution, as we will see now. Let us introduce the splitting of the ''initial data'' spaces onto an incoming and outgoing part. Technically, it can be performed in a straightforward way in the momentum representation. For each $1$-dimensional wave function defined on the original kinematical subspace $\Hil_{\nu}$ one can write the transformation
\begin{equation}
  [{\cal F}\Psi](b) = \sum_{\nu \in 4\mathbb{Z}^{+}} e^{i\frac{\nu b}{2}} \Psi(\nu) , 
\end{equation}
and the splitting
\begin{equation}\label{eq:split-trans}\begin{split}
  \Hil_{\nu} \ni |\Psi\rangle &\to |\Psi\rangle_{\pm} \in \Hil_{\nu} , \\
  \Psi_{\pm}(v) &= {\cal F}^{-1} \theta(\pm b) [{\cal F} \Psi](b) , 
\end{split}\end{equation}
where $\theta$ is the Heaviside step function. 

The same procedure can be easily applied to the eigenfunctions $e_{\vec{\omega}(\vec{k},h_{\vec{k}})}(\vec{\nu})$, using a straightforward extension of the formula \eqref{eq:split-trans}, since they must be treated as distributions rather than normalizable states on a Hilbert space. The corresponding eigenfunctions $e^{\pm}_{\vec{\omega},\vec{k}}$ are well defined and generically their value never reaches zero. This allows us to associate with each physical state \eqref{eq:phy-fin} the following initial data state
\begin{equation}\label{eq:prof-state}\begin{split}
  \Psi^{\pm}_{\vec{\nu}}(\vec{k}) 
  &= \sum_{\vec{k}\in(\mathbb{Z}^\star)^n} \int_0^{h_{\star}(\vec{k})} \rd h_{\vec{k}} \tilde{\Psi}(\vec{k},h_{\vec{k}}) 
		\frac{e^{\pm}_{\vec{\omega}(\vec{k},h_{\vec{k}})}(\vec{\nu})}{|e^{\pm}_{\vec{\omega}(\vec{k},h_{\vec{k}})}(\vec{\nu})|} ,
\end{split}\end{equation}
This association defines the desired unitary transformation $U_{\vec{\nu}}^{\pm}:\Hil_{{\rm phy}}\to \Hil_{{\rm phy}}^{\pm}$, and warranties on $\Hil_{{\rm phy}}^{\pm}$ the following inner product
\begin{equation}
	\langle\Phi_{\vec{\nu}} |\Psi_{\vec{\nu}} \rangle^{\pm}
	= \sum_{\vec{k}\in(\mathbb{Z}^\star)^n} \int_0^{h_{\star}(\vec{k})} \rd h_{\vec{k}} 
	\tilde{\Phi}^\star(\vec{k},h_{\vec{k}}) \tilde{\Psi}(\vec{k},h_{\vec{k}}) .
\end{equation}
The inner product, as expected, is time independent. Therefore, the evolution can be understood as the unitary transformation between the initial data spaces given by $U^{\pm}_{\vec{\nu}_1,\vec{\nu}_2} := U^{\pm}_{\vec{\nu}_2} (U^{\pm}_{\vec{\nu}_1})^{-1}$ whose action can be easily deduced from \eqref{eq:prof-state}. More explicitly,
\begin{equation}
	U^{\pm}_{\vec{\nu}} \tilde{\Psi}(\vec{k},h_{\vec{k}}) =\tilde{\Psi}(\vec{k},h_{\vec{k}}) 
	\frac{e^{\pm}_{\vec{\omega}(\vec{k},h_{\vec{k}})}(\vec{\nu})}{|e^{\pm}_{\vec{\omega}(\vec{k},h_{\vec{k}})}(\vec{\nu})|}
\end{equation}
Thus, given an observable $\hat O$ which can be represented as an operator acting directly on a spectral profile $\tilde{\Psi}(\vec{k},h_{\vec{k}})$ one can construct a family of the evolving observables in a straightforward way: 
\begin{equation}
  \hat O^{\pm}_{\vec{\nu}}=(U^{\pm}_{\vec{\nu}})^{-1}\hat O U^{\pm}_{\vec{\nu}}.
\end{equation}
These observables will be related by the transformation
\begin{equation}
	\hat O^{\pm}_{\vec{\nu}_2}=U^{\pm}_{\vec{\nu}_1,\vec{\nu}_2}\hat O^{\pm}_{\vec{\nu}_1}U^{\pm}_{\vec{\nu}_2,\vec{\nu}_1}.
\end{equation}

The choice of the evolution parameter is of course not restricted to just the two cases discussed above. In principle one can select for that purpose any function of the phase space variables. In particular in the context of black holes the variable $K_x$ being mathematical analog of $b$ (in the quantization scheme applied there) has been considered in \cite{BH-1,BH-2}. On the other hand in the present formulation of the model the variable $K_x$, while being a promising candidate for an internal time from classical considerations, it is a nontrivial function of \emph{both} $\nu_j$ and $b_j$. Thus the identification of the equivalents of the initial data spaces (like $\Hil^{\pm}_{\vec{\nu}}$) or the unitary transformations analogous to $U^{\pm}_{\vec{\nu}_1,\vec{\nu}_2}$ is much more involved and a nontrivial extension of the functional analysis techniques will be required to implement it.

As a final remark, it is worth to mention that, following the original construction in spherically symmetric spacetimes, we assume that we are able to distinguish the labels of vertices/edges of the graph supporting the reduced spin network. This assumption allows us to build observables indexed by the index $j$ (associated to a vertex/edge). However, let us notice that here the graph is a chain embedded in $S^1$. Then, cyclic permutations of these indices are actually a symmetry of the graph. In this respect, and following the orthodox treatment of LQG, this symmetry should be averaged over during the construction of the observables. As a consequence, it is meaningless to define observables indexed just by $j$ as the averaging would remove the absolute labeling $j$ of vertices/edges. Only relative labellings are meaningful. In this sense, one should consider instead observables measuring the correlations between quantities at different vertices. For example, observables of the form $(\hat{A}_{j} - \hat{A}_{j'})^2$ with $j\neq j'$, after a suitable averaging with respect to the above-mentioned symmetries, would yield nontrivial expectation values. Another example is the Hamiltonian constraint itself, as specified in Eq. \eqref{eq:quant-scalar-constrh0}, which, by its very definition, would not be affected by this averaging.

\subsection{Semiclassical sector}

With the physical Hilbert space and the (construction method for) set of physically meaningful observables at our disposal we can embark on the task of probing the dynamical properties of our model. A particular point of interest is the behavior of the semiclassical states. Probing it in the asymptotic future/past (where the spacetime is expected to reach its low curvature regime) is especially important in order to verify whether the constructed quantum model reproduces (the appropriate sector of) general relativity at low curvatures. Due to the inherent discreteness of the polymer quantization (here reflected in the spectra of the operators $\hat{\cal E}$ and $\hat{V}$), and the nonperturbative nature of the theory, the answer to this question is far from trivial.

The part of the semiclassical sector being of interest to our studies is distinguished by the necessary condition that the state is sharply peaked in the Dirac observables introduced in the previous subsection as well as in some appropriately selected family of evolving observables for some interval of the evolution parameter. Another, quite obvious necessary condition for these states is to approximate smooth manifolds. This amounts to the selection of states supported on the graphs with large number of vertices and for which the differences of expectation values of the relevant ``local'' observables (associated to a given vertex) between the consecutive vertices are small.
These states are particularly relevant in the understanding of the relation of the presented approach with nonsingular (bouncing) scenarios studied in LQC \cite{lqc}. Such comparison however is out of the scope of this manuscript and it will be left for future research. 

Relevant information can however be extracted in the context of the large volume limit (expected to correspond to the low energy limit) of the theory. Indeed, since (similarly to isotropic LQC) the eigenfunctions of $\hat{h}_j$ admit a well defined large $\nu_j$ limit---an analog of standing waves---one can use the scattering description of \cite{kp-scatter}. It is based on the observation (true for a wide range of systems studied in the LQC framework) that  at large volumes the eigenstates of either the Hamiltonian (or the evolution operator playing the role of it) or the Hamiltonian constraint converge on the one hand to certain combination of the analogous eigenstates in the standard geometrodynamics framework and on the other hand to simple analytic functions which encode the main physical properties of the system in the large volume regime. Those combinations represent either a standing or reflected waves (depending on the particular system). For the case studied in this article the asymptotics has been studied in detail in Appendix~\ref{app:disp-rel}. The dominant terms are combinations of exponentials $\exp[\pm i\sigma(\vec{k},h_{\vec{k}})\nu]$ (with equal amplitudes for $+$ and $-$ sign), where $\sigma(\vec{k},h_{\vec{k}})$ is a function of $\omega(\vec{k},h_{\vec{k}})$ specified via Eq. \eqref{eq:sigma-form}. Thus, the asymptotic future/past state of some spectral profile $\tilde{\Psi}^{\pm}(\vec{k},h_{\vec{k}})$  is characterized by
\begin{equation}\label{eq:asympt}
 \Psi(\vec{k},\vec{\nu}) 
	= \sum_{\vec{k}\in (\mathbb{Z}^\star)^n} \int_0^{h_\star(\vec{k})} \rd h_{\vec{k}} \tilde{\Psi}^{\pm}(\vec{k},h_{\vec{k}})
	e^{\pm i\sigma(\vec{k},h_{\vec{k}})\nu} .
\end{equation}
The asymptotic spectral profiles $\tilde{\Psi}^{\pm}(\vec{k},h_{\vec{k}})$ are determined by the original spectral profile of the 
complete state $\tilde{\Psi}(\vec{k},h_{\vec{k}})$, however the relation is a nontrivial $(\vec{k},h)$-dependent phase rotation
which usually has to be determined numerically (see for example \cite{kp-scatter,ppwe-rad,p-coh}).

One qualitative observation one can make immediately without analyzing the relation between the asymptotic states and the exact one is the counting of free semiclassical degrees of freedom emerging from the treatment. 
Since for the asymptotic spectral profile one can freely choose any function defined on an appropriate domain within $(\mathbb{Z}^\star)^n\times\mathbb{R}$ normalizable in the scalar product \eqref{eq:scalar-prod}, for instance, one can consider Gaussians sharply peaked about any sequence of $\vec{k}$ and $h_{\vec{k}}$. Furthermore by setting the appropriate rotations to these Gaussians one can arbitrarily shift the state in $\vec{\nu}$. As a consequence the state becomes truly ultralocal, as the peak can be set independently for each node. Namely, the variable conjugate to  $\vec{k}$  can take any arbitrary value.
Classically, the LRS Gowdy spacetime is completely characterized by a global degree of freedom, whereas its description in terms of the reduced Ashtekar--Barbero variables features the local unphysical degrees of freedom associated with the freedom of diffeomorphism transformations. There however the diffeomorphism constraint ties the data in distinct points of the reduced manifold by spatial derivatives. The quantization procedure implemented here removes this feature.

In consequence the space of solutions (and consequently of quantum trajectories) is much larger than that of GR. Thus, in the present form of the model, GR does not emerge solely as the large volume  limit of the (loop) quantum description. This excessive freedom can be traced back to the conjunction  of the procedure of Abelianization (making the Hamiltonian constraint ultralocal) and the qualitative differences in the treatment of Hamiltonian and diffeomorphism constraints. Since the original Hamiltonian constraint relates the quantum data at distinct vertices of the graph, in order to recover the correct count of the degrees of freedom one may be forced to implement the  diffeomorphism constraint in the same footing as the Hamiltonian one, that is by building the quantum counterpart of the regularized infinitesimal diffeomorphism generator \cite{mad}. This would 
provide the additional operator constraint, now mixing the data on distinct vertices of the graph.

\section{Conclusions}
\label{sec:conc}

To summarize, we have carried out a full quantization (within the LQG framework) of the polarized LRS Gowdy model in vacuum with $T^3$ topology. In the process no gauge fixing was implemented---the treatment remains diffeomorphism invariant. Our strategy is based on a suitable redefinition of the Hamiltonian constraint in such a way that it commutes with itself on both at the classical (under Poisson brackets) and at the quantum level. The resulting ultralocality of the Hamiltonian constraint allows then to find the solutions to it that are invariant under spacetime diffeomorphisms and to construct the physical Hilbert space. The observables of the model correspond to a global degree of freedom as in the classical theory and a new observable without classical Dirac observable analogue codifying the areas of the consecutive Killing orbits. A similar observable has been already identified  in spherically symmetric loop gravity \cite{BH-1}. The treatment allows us to probe the dynamics in an unambiguous way and the system admits a large semiclassical sector. A remarkable property of the dynamics is the singularity resolution with a mechanism similar to the one observed in LQC \cite{aps,hybrid}. The preliminary analysis of the asymptotic future/past epoch of the states suggests the necessity of either implementing the infinitesimal diffeomorphism constraint or finding alternative mechanisms preventing ultralocality. These results will on the one hand allow to verify the existing LQC frameworks against the genuine quantum nonperturbative dynamics of the inhomogeneous model and on the other hand provide a crucial information for the programs of probing the dynamical sector in full LQG, indicating new possible avenues for improving/completing the existing treatments.

Our results open new ways for the study of quantum gravity phenomenology in cosmology. On the one hand, they do not contradict the results of Ref. \cite{bojo-bra} for Gowdy cosmologies or Ref. \cite{bojo-bra-rey} in spherically symmetric gravity, regarding the difficulties for the avoidance of anomalies in the constraint algebra. On the other hand, it is worth commenting that the results of Refs. \cite{bojo-bra,bojo-bra-rey} do not exhaust all possible polymerizations of the Hamiltonian constraint and more general choices can solve the problem of the anomalies in the constraint algebra (see for instance Ref. \cite{cgop} for a counter example in nonvacuum spherically symmetric models). As we mentioned above, if we couple a scalar field to the full polarized Gowdy model, it is well known that the classical spacetime admits isotropic solutions with nonperturbative tensor and scalar fields propagating on them. A full quantization of this model in loop quantum gravity would allow us to check the validity of the hybrid quantization. For instance, if we identify  a semiclassical sector where there is a well defined notion of background geometry plus fields propagating on it, it would be interesting to probe the effective equations of motion and compare with the classical theory as well as with the hybrid quantization approach \cite{hybrid}. Besides, in the limit where the latter does not back react considerably with the background, it can have important consequences for the study of  tensor modes propagating in loop quantized Friedmann--Robertson--Walker spacetimes. For instance, it would be interesting to analyze the validity of the effective equations of motion considered presently in order to confront the predictions with observations. Not only if new corrections must be incorporated, but also if an effective (semiclassical) description is also valid in the deep quantum regime. If this is not the case, the present considerations about giving initial data at the bounce might be revisited. 

In addition, the quantum dynamics of the present LRS Gowdy model will be compared soon with the loop quantization of a LRS Bianchi I model. The results could seed light in the present understanding  of loop quantum cosmology and the different dynamical schemes that have been considered so far. 

\acknowledgments

We wish to thank R. Gambini, J. Lewandowski, G. A. Mena Marug\'an, J. Pullin, H. Sahlmann and T. Thiemann for comments. J. O. acknowledges the partial support by Pedeciba, Project No. NSF-PHY-1305000, Project No. PHY-1505411 and the Eberly research funds of Penn State University (USA). T. P. thanks the Polish Narodowe Centrum Nauki (NCN) grant 2012/05/E/ST2/03308. D. M.-dB. is supported by the project CONICYT/FONDECYT/POSTDOCTORADO/3140409 from Chile. The authors thank the grant MICINN FIS2014-54800-C2-2-P from Spain.

\appendix

\section{Alternative Abelianization procedure}\label{app:altern}

In the Abelianization procedure specified in sec.~\ref{sec:abelian}, when selecting the modified constraint algebra, we have conveniently multiplied the Hamiltonian constraint by the factor $\partial_{\theta}{\cal E}$. In this way the homogeneous sector of the theory can be analyzed classically, but at the price of extending (fortunately in a controlled way) the constraint surface with respect to the one in GR. This choice also introduces severe restrictions on the classical time evolution generated by the new Hamiltonian constraint, if one assumes that the new lapse function is well behaved.

For the sake of completeness, we also consider here a different scaling for the new Hamiltonian constraint. Instead of imposing regularity of the new constrained classical system in a neighborhood of the homogeneous sector, we would like to study the case in which the old Hamiltonian constraint enters in the definition of the new one with a factor one. Thus, the lapse function remains unchanged. Nonetheless, here, for simplicity, we will study the case in which the old Hamiltonian enters multiplied by a factor $(E^{x})^{-1}$. It is easy to realize that this constraint exhibits the same behavior than the former, since $(E^{x})^{-1}$ is a well behaved function (being in particular always finite and nowhere vanishing outside of classical singularity). Together with the remaining conditions for the transformation as specified in sec~\ref{sec:abelian} we end up replacing \eqref{eq:abel2} with \eqref{eq:abel1} multiplied by $(E^{x})^{-1}$. Unfortunately, this transformation is singular in the classical theory since the derivative $\partial_{\theta}{\cal E}$, on $S^1$, must vanish at least in two points. Thus, the studies following from it have to be treated carefully. We consider this choice only as an attempt to extend the range of the classical evolution with respect to the choice adopted in the main text of this manuscript. While on the phase space subset corresponding to slices where $\partial_{\theta}{\cal E}$ is generically nonzero, differentiability of the metric tensor is sufficient to ensure the preservation of the (appropriate portion of the) constraint surfaces, the points of the phase space (geometries) for which $\partial_{\theta}{\cal E}=0$ at some open set are removed! This may in particular break the spatial diffeomorphism group which may have an effect on the procedure of defining the diffeomorphism-invariant sector of the theory performed in sec.~\ref{sec:gave-diff}. 

Bearing in mind the caveats listed above we can now repeat the quantization procedure described in sec~\ref{sec:kin} to~\ref{sec:phys}. The kinematical quantization remains unchanged. Also, the construction of the Hamiltonian constraint in sec.~\ref{sec:qham} can be repeated directly. The resulting quantum Hamiltonian constraint takes the form very similar to \eqref{eq:quant-scalar-constrh0}
\begin{equation}
  \hat{\tilde{H}}(N)= \frac{\pi}{2\gamma\ell_{\rm Pl}}\sum_{j} \frac{N_j}{k_j-k_{j-1}}\hat P 
		\left[a_{j-1}^{3/2}\hat{h}_{j-1}-a_j^{3/2}\hat{h}_{j}\right]\hat P,
\end{equation}
where all the involved operators have already been defined in sec.~\ref{sec:qham}. This operator is not well defined on the subset ${\cal S}$ of the domain of $\tilde{H}$ defined via \eqref{eq:quant-scalar-constrh0} that contains $|\vec{k},\vec{\nu}\rangle$ for which any two consecutive components of $\vec{k}$ are equal. Due to the discreteness of $\Hil_{{\rm kin}}$ it is not obvious how this singularity cannot be circumvented.

With the Hamiltonian constraint operator at our disposal, we can now perform the averaging over the spin network embedding transformations which represent the subgroup of the finite spatial diffeomorphism group surviving after removing from $\Hil_{{\rm kin}}$ the subset ${\cal S}$. This averaging procedure, described in sec.~\ref{sec:gave-diff}, simply removes the embedding data while it does not introduce any restrictions on the quantum labels. In this way we define the analog $\tilde{\Hil}_{{\rm diff}}$ of the diffeomorphism invariant Hilbert space and the invariant Hamiltonian constraint operator acting on it. One has to remember though that the full diffeomorphism invariance is broken in our treatment and the invariant sector is not truly a diffeomorphism invariant sector of (a midisuperspace version of) LQG.

Since the invariant operator differs from the one in our original treatment only by factors $\Delta k_j := k_j-k_{j-1}$ one can directly repeat the derivation if its spectral decomposition and in consequence perform a group averaging over time reparametrizations as described in sec.~\ref{sec:qham}. As a result, the physical states are of the form\begin{equation}\label{eq:phy-fin-alt}
  \langle \vec{k}',\vec{\nu}' | \Psi \rangle 
	= \sum_{\vec{k}\in (\mathbb{Z}^\star)^n\setminus S} \int_0^{h_\star(\vec{k})} \rd h_{\vec{k}} \tilde{\Psi}(\vec{k},h_{\vec{k}})
	(e_{\vec{\omega}(\vec{k},h_{\vec{k}}),\vec{k}}|\vec{k}',\vec{\nu}'\rangle^\star
\end{equation}
where $S := \{ \vec{k}\in(\mathbb{Z}^\star)^n: \exists j\in \{2,\ldots,n\}: k_j = k_{j-1} \}$, 
and the physical inner product (induced from $\tilde{\Hil}_{\diff}$) is 
\begin{equation}
  \langle \Phi | \Psi \rangle = \sum_{\vec{k}\in (\mathbb{Z}^\star)^n\setminus S} \int_0^{h_\star(\vec{k})} \rd h_{\vec{k}}
    \tilde{\Phi}^\star(\vec{k},h_{\vec{k}}) \tilde{\Psi}(\vec{k},h_{\vec{k}}) .
\end{equation}
The construction of observables specified in sec.~\ref{sec:classLRS} can be then repeated directly.

As final comment we note that the space of solutions is smaller than the one resulting in sec. \ref{sec:classLRS} since the sectors ${\cal S}$ have been excluded. Nevertheless, it is not difficult to convince oneself about the existence of semiclassical sectors providing effective geometries with $\partial_{\theta}{\cal E}=0$ up to small (Planck order) corrections.

\section{Spectral properties of the Hamiltonian component operators}\label{app:disp-rel}

In this appendix we will describe some of the main qualitative results of probing the spectrum of the difference operators present in our analysis. In particular, we will study in detail the family of operators $\tilde{h}_{m,n}$ defined in \eqref{eq:hred} that are the basic component of the scalar constraint. This operators have a well defined action on the domain $\mathcal{D}$ of finite combinations of orthonormal basis elements $|\nu\rangle$ in the space $L^2(\bar{\re}_{\rm Bohr},\rd\mu_{\rm Bohr})$. That action reads
\begin{equation}\begin{split}
  \hat h_{m,n}|\nu\rangle&=g_+(\nu)|\nu+4\rangle+g_-(\nu)|\nu-4\rangle\\
  &+g_0(\nu)|\nu\rangle,
\end{split}\end{equation}
where
\begin{subequations} \begin{align}
  \begin{split}
    g_\pm(\nu)&=-\frac{1}{8(\gamma\lambda)^2}s_{\pm}(\nu)s_{\pm}(\nu\pm 2)b(\nu)b(\nu\pm4)\\
    &\times|\nu|^{1/4}|\nu\pm4|^{1/4}|\nu\pm 2|^{1/2},
  \end{split} \\
  \begin{split}
    g_0(\nu)&=\frac{1}{8(\gamma\lambda)^2} \left(b(\nu)^2|\nu|^{1/2}|\nu+2|^{1/2}s_+^2(\nu)\right. \\
    &\left.+b(\nu)^2|\nu|^{1/2}|\nu-2|^{1/2}s_-^2(\nu)\right)\ \\
    &-\frac{1}{2\lambda^2}b^4(\nu)(m-n)^2,
  \end{split}
\end{align}\end{subequations}
and $s_\pm(\nu)={\rm sgn}(\nu)+{\rm sgn}(\nu\pm 2)$. 

Fortunately for us, the eigenvalue problem  $\tilde{h}_{m,n}|\omega_j\rangle=\omega_j|\omega_j\rangle$ relevant for the goal specified above can be (at least in part) analyzed analytically, without having to rely on numerical tools. We first notice that any solution to this equation takes values on semilattices in the variable $\nu$ of step four. Therefore, they can be labeled by $\nu(\epsilon,l)=\epsilon+4l$, with $\epsilon_j\in(0,4]$ and $l\in \mathbb{N}_0$. Besides, these solutions are unique up to a normalization condition. This means that we only need to provide the initial data at the section $\nu=\epsilon$. As a consequence all the eigenspaces are of dimension one. In particular, the spectrum of $\tilde{h}_{m,n}$ is nondegenerate. 

To further determine the properties of the solutions to the eigenvalue problem we follow the ideas of Ref.~\cite{kp-scatter} already employed in isotropic spacetimes in loop quantum cosmology. In this case, we need to write the eigenvalue equation in a matrix form as follows. Let us define $e_{\omega}(\nu)=\langle\nu|\omega\rangle$ and introduce the vector
\begin{equation}
	{\bf e}_{\omega}(\nu)=\left(
	\begin{array}{c}
		e_{\omega}(\nu)\\
		e_{\omega}(\nu-4)
	\end{array}
	\right).
\end{equation}
Then, the eigenvalue equation can be written as 
\begin{equation}\label{eq:eig-1d}
	{\bf e}_{\omega}(\nu+4)={\bf A}_{\omega}(\nu){\bf e}_{\omega}(\nu),
\end{equation}
where
\begin{equation}
	{\bf A}_{\omega}(\nu)=\left(
	\begin{array}{cc}
		\frac{\omega-g_{0}(\nu)}{g_-(\nu+4)}&-\frac{g_+(\nu-4)}{g_-(\nu+4)}\\
		1&0
	\end{array}
	\right).
\end{equation}
In the next step we transform the eigenvalue equation \eqref{eq:eig-1d} as the equation involving the coefficients of the decompositions of ${\bf e}_{\omega}(\nu)$ with respect to the functions $\underline{e}_{\varkappa}(\nu)$, which are selected to be of a simple analytic form and are expected to well approximate the behavior of the eigenfunction for large $\nu$. The particular form of these functions is guessed from the form of the exact eigenfunctions for large $\nu$ determined either numerical or analytical analysis. In our case the natural choice is
\begin{equation}\label{eq:lim-prop}
  \underline{e}_{\varkappa}^{\pm}(\nu) = \exp(\pm \varkappa \nu) , \qquad \varkappa \in \mathbb{C} .
\end{equation}
To perform the transformation let us define the matrix 
\begin{equation}
  {\bf B}_{\varkappa}(\nu)=\left(
  \begin{array}{cc}
    \underline{e}_{\varkappa}^{+}(\nu+4)&\underline{e}^{-}_{\varkappa}(\nu+4)\\
    \underline{e}_{\varkappa}^{+}(\nu)&\underline{e}^{-}_{\varkappa}(\nu)
  \end{array}
  \right).
\end{equation}
The equation \eqref{eq:eig-1d} can then be written in the form ${\bf e}_{\omega}(\nu)={\bf B}_{\varkappa}(\nu)\tilde{{\bf e}}_{\omega}(\nu)$ where the new ``eigenfunction coefficients'' satisfy the equation
\begin{equation}\begin{split}
  \tilde{{\bf e}}_{\omega}(\nu+4)
  &={\bf B}^{-1}_{\varkappa}(\nu){\bf A}_{\omega}(\nu){\bf B}_{\varkappa}(\nu-4)\tilde{{\bf e}}_{\omega}(\nu)  \\
  &={\bf M}_{\omega}(\nu)\tilde{{\bf e}}_{\omega}(\nu).
\end{split}\end{equation}
Now, for the eigenfunction to actually converge in the large $\nu$ limit to a certain combination of $\underline{e}_{\varkappa}^{\pm}$, the matrices ${\bf M}_{\omega}(\nu)$ need to converge in that limit to the unity, namely 
\begin{equation}\label{eq:M-pre} 
  \lim_{\nu\to\infty}{\bf M}_{\omega}(\nu) = \mathbf{I} .  
\end{equation}
By inspection this condition is satisfied if and only if the following relation between $\omega$ and $\varkappa$ (the dispersion relation) holds
\begin{equation}\label{eq:disp-gen}
  1 - \cosh(\varkappa) = (\gamma\lambda)^2 \omega . 
\end{equation}
The condition \eqref{eq:M-pre} is not sufficient for the considered convergence of the eigenfunctions to hold. For that, ${\bf M}_{\omega_j}(\nu_j)$ must approach the unity sufficiently fast. However, by direct inspection we see that, provided the dispersion relation \eqref{eq:disp-gen} holds, we actually have
\begin{equation}
  {\bf M}_{\omega}(\nu) = \mathbf{I} + O(\nu^{-2}) ,
\end{equation}
which is sufficient to ensure the required explicit convergence.

Once we have verified that the combinations of the functions proposed in \eqref{eq:lim-prop} indeed provide a suitable large $\nu$ limit of the eigenfunctions of $\tilde{h}_{m,n}$, we can apply them to determine the properties of these operators, in particular to verify their self-adjointness through the analysis of the deficiency subspaces and to probe their spectra.

\subsection{Self-adjointness}

Let us start with the analysis of the deficiency subspaces of the operators $\tilde{h}_{m,n}$. By direct inspection, one can see that all of them are symmetric. Their deficiency subspaces ${\cal K}^{\pm}_{m,n}$ are the spaces of normalizable solutions $\psi^{\pm}_{m,n}$ to the equation
\begin{equation}
  (\psi^{\pm}_{m,n}| \tilde{h}_{m,n} \mp i \id |\chi\rangle = 0,\quad \forall |\chi\rangle \in \mathcal{D}. 
\end{equation}
By the nondegeneracy of the eigenvalue problem these spaces are at most $1$-dimensional. To verify whether the solutions $\psi^{\pm}_{m,n}$ are normalizable we inspect their large $\nu$ limit. It is given by a combination of the functions \eqref{eq:lim-prop} for $\omega = \pm i$ 
\begin{equation}
  \psi^{\pm}_{m,n}(\nu) = [c^{\pm}_{+,m,n} + O(\nu^{-2})] e^{\varkappa_{\pm} \nu} 
  + [c^{\pm}_{-,m,n} + O(\nu^{-2})] e^{-\varkappa_{\pm} \nu} ,
\end{equation}
where $\varkappa_{\pm}$ is the solution to \eqref{eq:disp-gen} for $\omega = \pm i$. It can be decomposed into its real 
and imaginary parts $\varkappa_{\pm}=:\rho_{\pm}+i\sigma_{\pm}$. Then the equation \eqref{eq:disp-gen} reads
\begin{subequations}\label{eq:def-k-dec}\begin{align}
  \cosh(\rho_\pm)\cos(\sigma_\pm) &= 1 , \\
  \sinh(\rho_\pm)\sin(\sigma_\pm) &= \pm (\gamma\lambda)^2 .
\end{align}\end{subequations}
Since for our choice of the limit basis \eqref{eq:lim-prop} we have 
$\underline{e}_{-\varkappa}^{\pm}(\nu) = \underline{e}_{\varkappa}^{\mp}(\nu)$ the sign of $\rho_{\pm}$ can be fixed without 
the loss of generality. We choose it to be $\rho_{\pm}>0$. 

By direct inspection of the system \eqref{eq:def-k-dec} we notice that the limit of any deficiency function will have 
two components: one exponentially growing and one exponentially decaying. Thus, for these functions to be normalizable, the 
coefficients $c^+_{+,m,n}$ and $c^-_{-,m,n}$ have to vanish. Since all the coefficients $c^{\pm}_{\pm,m,n}$ are continuous 
non-constant functions of $\gamma$, the set of values of $\gamma$ admitting $c^+_{+,m,n} = c^-_{-,m,n} = 0$ for at least one pair 
$m,n$ is at most nongeneric. It is worth commenting that since most of the approaches in LQG providing concrete values of $\gamma$ allow to find its value only numerically, no definite 
statement regarding nonvanishing of $c^\pm_{\pm,m,n}$ can be given at this point. However, the existence of normalizable deficiency 
functions would require an extreme fine tuning of this parameter. We can thus assume with great reliability that the deficiency 
spaces are trivial. Therefore, all the operators $\tilde{h}_{m,n}$ are essentially self-adjoint, although the exact formal proof is 
not complete.

\subsection{The spectra}

Once we have established the (essential) self-adjointness of the studied operators we can focus on their eigenspaces corresponding to real eigenvalues and determine their spectra. 

When $\omega$ is restricted to real values only we notice that, if $\varkappa = \rho+i\sigma$, then
\begin{subequations}\label{eq:def-k}\begin{align}
  \cosh(\rho)\cos(\sigma) &= 1-(\gamma\lambda)^2\omega, \\
  \sinh(\rho)\sin(\sigma) &= 0.
\end{align}\end{subequations}
We observe two regimes with qualitatively distinct behavior of the limit of the eigenfunctions
\begin{enumerate}
	\item For $\omega\in [0,2(\gamma\lambda)^{-2}]$ we have that $\varkappa = i\sigma$, i.e., it is purely imaginary and the  
		basis asymptotically 	consists of plane waves 
		on a lattice. Thus this interval is a continuous part of the spectrum and the actual dispersion relation takes the  form
		\begin{equation}\label{eq:sigma-form}
			1 - \cos(\sigma) = (\gamma\lambda)^{2}\omega . 
		\end{equation}
		Let us notice that it is the usual one for a particle on a lattice. Besides, no solutions can be found for $\omega>2(\gamma\lambda)^{-2}$ if $\omega\in\mathbb{R}^+$.
	\item For $\omega < 0$ the parameter $\varkappa=\rho$ is real and the basis \eqref{eq:lim-prop} asymptotically consists of 
		exponential functions, thus this set may contain the discrete part of the spectrum only. In this case, a more careful analysis is necessary. We do not carry it out here since we are interested in the semiclassical sector already codified in the continuum part of the spectrum.
\end{enumerate}

\bibliography{gowdybib}{}

\begin{thebibliography}{70}%
\makeatletter
\providecommand \@ifxundefined [1]{%
 \@ifx{#1\undefined}
}%
\providecommand \@ifnum [1]{%
 \ifnum #1\expandafter \@firstoftwo
 \else \expandafter \@secondoftwo
 \fi
}%
\providecommand \@ifx [1]{%
 \ifx #1\expandafter \@firstoftwo
 \else \expandafter \@secondoftwo
 \fi
}%
\providecommand \natexlab [1]{#1}%
\providecommand \enquote  [1]{``#1''}%
\providecommand \bibnamefont  [1]{#1}%
\providecommand \bibfnamefont [1]{#1}%
\providecommand \citenamefont [1]{#1}%
\providecommand \href@noop [0]{\@secondoftwo}%
\providecommand \href [0]{\begingroup \@sanitize@url \@href}%
\providecommand \@href[1]{\@@startlink{#1}\@@href}%
\providecommand \@@href[1]{\endgroup#1\@@endlink}%
\providecommand \@sanitize@url [0]{\catcode `\\12\catcode `\$12\catcode
  `\&12\catcode `\#12\catcode `\^12\catcode `\_12\catcode `\%12\relax}%
\providecommand \@@startlink[1]{}%
\providecommand \@@endlink[0]{}%
\providecommand \url  [0]{\begingroup\@sanitize@url \@url }%
\providecommand \@url [1]{\endgroup\@href {#1}{\urlprefix }}%
\providecommand \urlprefix  [0]{URL }%
\providecommand \Eprint [0]{\href }%
\providecommand \doibase [0]{http://dx.doi.org/}%
\providecommand \selectlanguage [0]{\@gobble}%
\providecommand \bibinfo  [0]{\@secondoftwo}%
\providecommand \bibfield  [0]{\@secondoftwo}%
\providecommand \translation [1]{[#1]}%
\providecommand \BibitemOpen [0]{}%
\providecommand \bibitemStop [0]{}%
\providecommand \bibitemNoStop [0]{.\EOS\space}%
\providecommand \EOS [0]{\spacefactor3000\relax}%
\providecommand \BibitemShut  [1]{\csname bibitem#1\endcsname}%
\let\auto@bib@innerbib\@empty
\bibitem [{\citenamefont {Gowdy}(1971)}]{gowdy1}%
  \BibitemOpen
  \bibfield  {author} {\bibinfo {author} {\bibfnamefont {R.~H.}\ \bibnamefont
  {Gowdy}},\ }\href {\doibase 10.1103/PhysRevLett.27.826} {\bibfield  {journal}
  {\bibinfo  {journal} {Phys. Rev. Lett.}\ }\textbf {\bibinfo {volume} {27}},\
  \bibinfo {pages} {826} (\bibinfo {year} {1971})}\BibitemShut {NoStop}%
\bibitem [{\citenamefont {Bianchi}(1897)}]{bianchi}%
  \BibitemOpen
  \bibfield  {author} {\bibinfo {author} {\bibfnamefont {L.}~\bibnamefont
  {Bianchi}},\ }\href@noop {} {\bibfield  {journal} {\bibinfo  {journal} {Mem.
  Soc. It. Della Sc.(Dei XL)}\ }\textbf {\bibinfo {volume} {11}},\ \bibinfo
  {pages} {267} (\bibinfo {year} {1897})}\BibitemShut {NoStop}%
\bibitem [{\citenamefont {Misner}(1973)}]{misner}%
  \BibitemOpen
  \bibfield  {author} {\bibinfo {author} {\bibfnamefont {C.~W.}\ \bibnamefont
  {Misner}},\ }\href {\doibase 10.1103/PhysRevD.8.3271} {\bibfield  {journal}
  {\bibinfo  {journal} {Phys. Rev. D}\ }\textbf {\bibinfo {volume} {8}},\
  \bibinfo {pages} {3271} (\bibinfo {year} {1973})}\BibitemShut {NoStop}%
\bibitem [{\citenamefont {Berger}(1974)}]{berger1}%
  \BibitemOpen
  \bibfield  {author} {\bibinfo {author} {\bibfnamefont {B.}~\bibnamefont
  {Berger}},\ }\href@noop {} {\bibfield  {journal} {\bibinfo  {journal} {Ann.
  Phys.}\ }\textbf {\bibinfo {volume} {83}},\ \bibinfo {pages} {458} (\bibinfo
  {year} {1974})}\BibitemShut {NoStop}%
\bibitem [{\citenamefont {Pierri}(2002)}]{pierri}%
  \BibitemOpen
  \bibfield  {author} {\bibinfo {author} {\bibfnamefont {M.}~\bibnamefont
  {Pierri}},\ }\href {\doibase 10.1142/S0218271802001779} {\bibfield  {journal}
  {\bibinfo  {journal} {Int. J. Mod. Phys. D}\ }\textbf {\bibinfo {volume}
  {11}},\ \bibinfo {pages} {135} (\bibinfo {year} {2002})},\ \Eprint
  {http://arxiv.org/abs/gr-qc/0101013} {arXiv:gr-qc/0101013 [gr-qc]}
  \BibitemShut {NoStop}%
\bibitem [{\citenamefont {Corichi}\ \emph {et~al.}(2002)\citenamefont
  {Corichi}, \citenamefont {Cortez},\ and\ \citenamefont {Quevedo}}]{unit1}%
  \BibitemOpen
  \bibfield  {author} {\bibinfo {author} {\bibfnamefont {A.}~\bibnamefont
  {Corichi}}, \bibinfo {author} {\bibfnamefont {J.}~\bibnamefont {Cortez}}, \
  and\ \bibinfo {author} {\bibfnamefont {H.}~\bibnamefont {Quevedo}},\ }\href
  {\doibase 10.1142/S0218271802002281} {\bibfield  {journal} {\bibinfo
  {journal} {Int. J. Mod. Phys. D}\ }\textbf {\bibinfo {volume} {11}},\
  \bibinfo {pages} {1451} (\bibinfo {year} {2002})},\ \Eprint
  {http://arxiv.org/abs/gr-qc/0204053} {arXiv:gr-qc/0204053 [gr-qc]}
  \BibitemShut {NoStop}%
\bibitem [{\citenamefont {Cortez}\ and\ \citenamefont
  {Mena~Marugan}(2005)}]{unit2}%
  \BibitemOpen
  \bibfield  {author} {\bibinfo {author} {\bibfnamefont {J.}~\bibnamefont
  {Cortez}}\ and\ \bibinfo {author} {\bibfnamefont {G.~A.}\ \bibnamefont
  {Mena~Marugan}},\ }\href {\doibase 10.1103/PhysRevD.72.064020} {\bibfield
  {journal} {\bibinfo  {journal} {Phys. Rev. D}\ }\textbf {\bibinfo {volume}
  {72}},\ \bibinfo {pages} {064020} (\bibinfo {year} {2005})},\ \Eprint
  {http://arxiv.org/abs/gr-qc/0507139} {arXiv:gr-qc/0507139 [gr-qc]}
  \BibitemShut {NoStop}%
\bibitem [{\citenamefont {Corichi}\ \emph
  {et~al.}(2006{\natexlab{a}})\citenamefont {Corichi}, \citenamefont {Cortez},\
  and\ \citenamefont {Mena~Marugan}}]{unit3}%
  \BibitemOpen
  \bibfield  {author} {\bibinfo {author} {\bibfnamefont {A.}~\bibnamefont
  {Corichi}}, \bibinfo {author} {\bibfnamefont {J.}~\bibnamefont {Cortez}}, \
  and\ \bibinfo {author} {\bibfnamefont {G.~A.}\ \bibnamefont {Mena~Marugan}},\
  }\href {\doibase 10.1103/PhysRevD.73.084020} {\bibfield  {journal} {\bibinfo
  {journal} {Phys. Rev. D}\ }\textbf {\bibinfo {volume} {73}},\ \bibinfo
  {pages} {084020} (\bibinfo {year} {2006}{\natexlab{a}})},\ \Eprint
  {http://arxiv.org/abs/gr-qc/0603006} {arXiv:gr-qc/0603006 [gr-qc]}
  \BibitemShut {NoStop}%
\bibitem [{\citenamefont {Corichi}\ \emph
  {et~al.}(2006{\natexlab{b}})\citenamefont {Corichi}, \citenamefont {Cortez},\
  and\ \citenamefont {Mena~Marugan}}]{unit4}%
  \BibitemOpen
  \bibfield  {author} {\bibinfo {author} {\bibfnamefont {A.}~\bibnamefont
  {Corichi}}, \bibinfo {author} {\bibfnamefont {J.}~\bibnamefont {Cortez}}, \
  and\ \bibinfo {author} {\bibfnamefont {G.~A.}\ \bibnamefont {Mena~Marugan}},\
  }\href {\doibase 10.1103/PhysRevD.73.041502} {\bibfield  {journal} {\bibinfo
  {journal} {Phys. Rev. D}\ }\textbf {\bibinfo {volume} {73}},\ \bibinfo
  {pages} {041502} (\bibinfo {year} {2006}{\natexlab{b}})},\ \Eprint
  {http://arxiv.org/abs/gr-qc/0510109} {arXiv:gr-qc/0510109 [gr-qc]}
  \BibitemShut {NoStop}%
\bibitem [{\citenamefont {Corichi}\ \emph {et~al.}(2007)\citenamefont
  {Corichi}, \citenamefont {Cortez}, \citenamefont {Mena~Marugan},\ and\
  \citenamefont {Velhinho}}]{unit5}%
  \BibitemOpen
  \bibfield  {author} {\bibinfo {author} {\bibfnamefont {A.}~\bibnamefont
  {Corichi}}, \bibinfo {author} {\bibfnamefont {J.}~\bibnamefont {Cortez}},
  \bibinfo {author} {\bibfnamefont {G.~A.}\ \bibnamefont {Mena~Marugan}}, \
  and\ \bibinfo {author} {\bibfnamefont {J.~M.}\ \bibnamefont {Velhinho}},\
  }\href {\doibase 10.1103/PhysRevD.76.124031} {\bibfield  {journal} {\bibinfo
  {journal} {Phys. Rev. D}\ }\textbf {\bibinfo {volume} {76}},\ \bibinfo
  {pages} {124031} (\bibinfo {year} {2007})},\ \Eprint
  {http://arxiv.org/abs/0710.0277} {arXiv:0710.0277 [gr-qc]} \BibitemShut
  {NoStop}%
\bibitem [{\citenamefont {Corichi}\ \emph
  {et~al.}(2006{\natexlab{c}})\citenamefont {Corichi}, \citenamefont {Cortez},
  \citenamefont {Mena~Marugan},\ and\ \citenamefont {Velhinho}}]{uniq1}%
  \BibitemOpen
  \bibfield  {author} {\bibinfo {author} {\bibfnamefont {A.}~\bibnamefont
  {Corichi}}, \bibinfo {author} {\bibfnamefont {J.}~\bibnamefont {Cortez}},
  \bibinfo {author} {\bibfnamefont {G.~A.}\ \bibnamefont {Mena~Marugan}}, \
  and\ \bibinfo {author} {\bibfnamefont {J.~M.}\ \bibnamefont {Velhinho}},\
  }\href {\doibase 10.1088/0264-9381/23/22/014} {\bibfield  {journal} {\bibinfo
   {journal} {Class. Quant. Grav.}\ }\textbf {\bibinfo {volume} {23}},\
  \bibinfo {pages} {6301} (\bibinfo {year} {2006}{\natexlab{c}})},\ \Eprint
  {http://arxiv.org/abs/gr-qc/0607136} {arXiv:gr-qc/0607136 [gr-qc]}
  \BibitemShut {NoStop}%
\bibitem [{\citenamefont {Cortez}\ \emph {et~al.}(2007)\citenamefont {Cortez},
  \citenamefont {Mena~Marugan},\ and\ \citenamefont {Velhinho}}]{uniq2}%
  \BibitemOpen
  \bibfield  {author} {\bibinfo {author} {\bibfnamefont {J.}~\bibnamefont
  {Cortez}}, \bibinfo {author} {\bibfnamefont {G.~A.}\ \bibnamefont
  {Mena~Marugan}}, \ and\ \bibinfo {author} {\bibfnamefont {J.~M.}\
  \bibnamefont {Velhinho}},\ }\href {\doibase 10.1103/PhysRevD.75.084027}
  {\bibfield  {journal} {\bibinfo  {journal} {Phys. Rev. D}\ }\textbf {\bibinfo
  {volume} {75}},\ \bibinfo {pages} {084027} (\bibinfo {year} {2007})},\
  \Eprint {http://arxiv.org/abs/gr-qc/0702117} {arXiv:gr-qc/0702117 [GR-QC]}
  \BibitemShut {NoStop}%
\bibitem [{\citenamefont {Mena~Marugan}(1997)}]{mena1}%
  \BibitemOpen
  \bibfield  {author} {\bibinfo {author} {\bibfnamefont {G.~A.}\ \bibnamefont
  {Mena~Marugan}},\ }\href {\doibase 10.1103/PhysRevD.56.908} {\bibfield
  {journal} {\bibinfo  {journal} {Phys. Rev. D}\ }\textbf {\bibinfo {volume}
  {56}},\ \bibinfo {pages} {908} (\bibinfo {year} {1997})},\ \Eprint
  {http://arxiv.org/abs/gr-qc/9704041} {arXiv:gr-qc/9704041 [gr-qc]}
  \BibitemShut {NoStop}%
\bibitem [{\citenamefont {Banerjee}\ and\ \citenamefont
  {Date}(2008{\natexlab{a}})}]{baren}%
  \BibitemOpen
  \bibfield  {author} {\bibinfo {author} {\bibfnamefont {K.}~\bibnamefont
  {Banerjee}}\ and\ \bibinfo {author} {\bibfnamefont {G.}~\bibnamefont
  {Date}},\ }\href {\doibase 10.1088/0264-9381/25/10/105014} {\bibfield
  {journal} {\bibinfo  {journal} {Class. Quant. Grav.}\ }\textbf {\bibinfo
  {volume} {25}},\ \bibinfo {pages} {105014} (\bibinfo {year}
  {2008}{\natexlab{a}})},\ \Eprint {http://arxiv.org/abs/0712.0683}
  {arXiv:0712.0683 [gr-qc]} \BibitemShut {NoStop}%
\bibitem [{\citenamefont {Banerjee}\ and\ \citenamefont
  {Date}(2008{\natexlab{b}})}]{baren2}%
  \BibitemOpen
  \bibfield  {author} {\bibinfo {author} {\bibfnamefont {K.}~\bibnamefont
  {Banerjee}}\ and\ \bibinfo {author} {\bibfnamefont {G.}~\bibnamefont
  {Date}},\ }\href {\doibase 10.1088/0264-9381/25/14/145004} {\bibfield
  {journal} {\bibinfo  {journal} {Class. Quant. Grav.}\ }\textbf {\bibinfo
  {volume} {25}},\ \bibinfo {pages} {145004} (\bibinfo {year}
  {2008}{\natexlab{b}})},\ \Eprint {http://arxiv.org/abs/0712.0687}
  {arXiv:0712.0687 [gr-qc]} \BibitemShut {NoStop}%
\bibitem [{\citenamefont {Bojowald}\ and\ \citenamefont
  {Swiderski}(2004)}]{bs-spher}%
  \BibitemOpen
  \bibfield  {author} {\bibinfo {author} {\bibfnamefont {M.}~\bibnamefont
  {Bojowald}}\ and\ \bibinfo {author} {\bibfnamefont {R.}~\bibnamefont
  {Swiderski}},\ }\href {\doibase 10.1088/0264-9381/21/21/009} {\bibfield
  {journal} {\bibinfo  {journal} {Class. Quant. Grav.}\ }\textbf {\bibinfo
  {volume} {21}},\ \bibinfo {pages} {4881} (\bibinfo {year} {2004})},\ \Eprint
  {http://arxiv.org/abs/gr-qc/0407018} {arXiv:gr-qc/0407018 [gr-qc]}
  \BibitemShut {NoStop}%
\bibitem [{\citenamefont {Thiemann}(2008)}]{qsd}%
  \BibitemOpen
  \bibfield  {author} {\bibinfo {author} {\bibfnamefont {T.}~\bibnamefont
  {Thiemann}},\ }\href
  {http://www.cambridge.org/catalogue/catalogue.asp?isbn=9780521842631} {\emph
  {\bibinfo {title} {{Modern canonical quantum general relativity}}}}\
  (\bibinfo  {publisher} {Cambridge University Press},\ \bibinfo {year}
  {2008})\ \Eprint {http://arxiv.org/abs/gr-qc/0110034} {arXiv:gr-qc/0110034
  [gr-qc]} \BibitemShut {NoStop}%
\bibitem [{\citenamefont {Martin-Benito}\ \emph
  {et~al.}(2008{\natexlab{a}})\citenamefont {Martin-Benito}, \citenamefont
  {Garay},\ and\ \citenamefont {Mena~Marugan}}]{hybrid}%
  \BibitemOpen
  \bibfield  {author} {\bibinfo {author} {\bibfnamefont {M.}~\bibnamefont
  {Martin-Benito}}, \bibinfo {author} {\bibfnamefont {L.~J.}\ \bibnamefont
  {Garay}}, \ and\ \bibinfo {author} {\bibfnamefont {G.~A.}\ \bibnamefont
  {Mena~Marugan}},\ }\href {\doibase 10.1103/PhysRevD.78.083516} {\bibfield
  {journal} {\bibinfo  {journal} {Phys. Rev. D}\ }\textbf {\bibinfo {volume}
  {78}},\ \bibinfo {pages} {083516} (\bibinfo {year} {2008}{\natexlab{a}})},\
  \Eprint {http://arxiv.org/abs/0804.1098} {arXiv:0804.1098 [gr-qc]}
  \BibitemShut {NoStop}%
\bibitem [{\citenamefont {Mart\'in-Benito}\ \emph {et~al.}(2010)\citenamefont
  {Mart\'in-Benito}, \citenamefont {Marug\'an},\ and\ \citenamefont
  {Wilson-Ewing}}]{hybrid1}%
  \BibitemOpen
  \bibfield  {author} {\bibinfo {author} {\bibfnamefont {M.}~\bibnamefont
  {Mart\'in-Benito}}, \bibinfo {author} {\bibfnamefont {G.~A.~M.}\ \bibnamefont
  {Marug\'an}}, \ and\ \bibinfo {author} {\bibfnamefont {E.}~\bibnamefont
  {Wilson-Ewing}},\ }\href {\doibase 10.1103/PhysRevD.82.084012} {\bibfield
  {journal} {\bibinfo  {journal} {Phys. Rev. D}\ }\textbf {\bibinfo {volume}
  {82}},\ \bibinfo {pages} {084012} (\bibinfo {year} {2010})},\ \Eprint
  {http://arxiv.org/abs/1006.2369} {arXiv:1006.2369 [gr-qc]} \BibitemShut
  {NoStop}%
\bibitem [{\citenamefont {Garay}\ \emph {et~al.}(2010)\citenamefont {Garay},
  \citenamefont {Martin-Benito},\ and\ \citenamefont {Mena~Marugan}}]{hybrid2}%
  \BibitemOpen
  \bibfield  {author} {\bibinfo {author} {\bibfnamefont {L.~J.}\ \bibnamefont
  {Garay}}, \bibinfo {author} {\bibfnamefont {M.}~\bibnamefont
  {Martin-Benito}}, \ and\ \bibinfo {author} {\bibfnamefont {G.~A.}\
  \bibnamefont {Mena~Marugan}},\ }\href {\doibase 10.1103/PhysRevD.82.044048}
  {\bibfield  {journal} {\bibinfo  {journal} {Phys. Rev. D}\ }\textbf {\bibinfo
  {volume} {82}},\ \bibinfo {pages} {044048} (\bibinfo {year} {2010})},\
  \Eprint {http://arxiv.org/abs/1005.5654} {arXiv:1005.5654 [gr-qc]}
  \BibitemShut {NoStop}%
\bibitem [{\citenamefont {Fernandez-Mendez}\ \emph {et~al.}(2012)\citenamefont
  {Fernandez-Mendez}, \citenamefont {Marugan}, \citenamefont {Olmedo},\ and\
  \citenamefont {Velhinho}}]{hyb-inf}%
  \BibitemOpen
  \bibfield  {author} {\bibinfo {author} {\bibfnamefont {M.}~\bibnamefont
  {Fernandez-Mendez}}, \bibinfo {author} {\bibfnamefont {G.~A.~M.}\
  \bibnamefont {Marugan}}, \bibinfo {author} {\bibfnamefont {J.}~\bibnamefont
  {Olmedo}}, \ and\ \bibinfo {author} {\bibfnamefont {J.~M.}\ \bibnamefont
  {Velhinho}},\ }\href {\doibase 10.1103/PhysRevD.85.103525} {\bibfield
  {journal} {\bibinfo  {journal} {Phys. Rev. D}\ }\textbf {\bibinfo {volume}
  {85}},\ \bibinfo {pages} {103525} (\bibinfo {year} {2012})},\ \Eprint
  {http://arxiv.org/abs/1203.2525} {arXiv:1203.2525 [gr-qc]} \BibitemShut
  {NoStop}%
\bibitem [{\citenamefont {Fern\'andez-M\'endez}\ \emph
  {et~al.}(2012)\citenamefont {Fern\'andez-M\'endez}, \citenamefont
  {Mena~Marug\'an},\ and\ \citenamefont {Olmedo}}]{hyb-inf1}%
  \BibitemOpen
  \bibfield  {author} {\bibinfo {author} {\bibfnamefont {M.}~\bibnamefont
  {Fern\'andez-M\'endez}}, \bibinfo {author} {\bibfnamefont {G.~A.}\
  \bibnamefont {Mena~Marug\'an}}, \ and\ \bibinfo {author} {\bibfnamefont
  {J.}~\bibnamefont {Olmedo}},\ }\href {\doibase 10.1103/PhysRevD.86.024003}
  {\bibfield  {journal} {\bibinfo  {journal} {Phys. Rev. D}\ }\textbf {\bibinfo
  {volume} {86}},\ \bibinfo {pages} {024003} (\bibinfo {year} {2012})},\
  \Eprint {http://arxiv.org/abs/1205.1917} {arXiv:1205.1917 [gr-qc]}
  \BibitemShut {NoStop}%
\bibitem [{\citenamefont {Fern\'andez-M\'endez}\ \emph
  {et~al.}(2013)\citenamefont {Fern\'andez-M\'endez}, \citenamefont
  {Mena~Marug\'an},\ and\ \citenamefont {Olmedo}}]{hyb-inf2}%
  \BibitemOpen
  \bibfield  {author} {\bibinfo {author} {\bibfnamefont {M.}~\bibnamefont
  {Fern\'andez-M\'endez}}, \bibinfo {author} {\bibfnamefont {G.~A.}\
  \bibnamefont {Mena~Marug\'an}}, \ and\ \bibinfo {author} {\bibfnamefont
  {J.}~\bibnamefont {Olmedo}},\ }\href {\doibase 10.1103/PhysRevD.88.044013}
  {\bibfield  {journal} {\bibinfo  {journal} {Phys. Rev. D}\ }\textbf {\bibinfo
  {volume} {88}},\ \bibinfo {pages} {044013} (\bibinfo {year} {2013})},\
  \Eprint {http://arxiv.org/abs/1307.5222} {arXiv:1307.5222 [gr-qc]}
  \BibitemShut {NoStop}%
\bibitem [{\citenamefont {Fern\'andez-M\'endez}\ \emph
  {et~al.}(2014)\citenamefont {Fern\'andez-M\'endez}, \citenamefont
  {Mena~Marug\'an},\ and\ \citenamefont {Olmedo}}]{hyb-inf3}%
  \BibitemOpen
  \bibfield  {author} {\bibinfo {author} {\bibfnamefont {M.}~\bibnamefont
  {Fern\'andez-M\'endez}}, \bibinfo {author} {\bibfnamefont {G.~A.}\
  \bibnamefont {Mena~Marug\'an}}, \ and\ \bibinfo {author} {\bibfnamefont
  {J.}~\bibnamefont {Olmedo}},\ }\href {\doibase 10.1103/PhysRevD.89.044041}
  {\bibfield  {journal} {\bibinfo  {journal} {Phys. Rev. D}\ }\textbf {\bibinfo
  {volume} {89}},\ \bibinfo {pages} {044041} (\bibinfo {year} {2014})},\
  \Eprint {http://arxiv.org/abs/1401.5256} {arXiv:1401.5256 [gr-qc]}
  \BibitemShut {NoStop}%
\bibitem [{\citenamefont {Gomar}\ \emph {et~al.}(2014)\citenamefont {Gomar},
  \citenamefont {Fern\'andez-M\'endez}, \citenamefont {Marug\'an},\ and\
  \citenamefont {Olmedo}}]{hyb-inf4}%
  \BibitemOpen
  \bibfield  {author} {\bibinfo {author} {\bibfnamefont {L.~C.}\ \bibnamefont
  {Gomar}}, \bibinfo {author} {\bibfnamefont {M.}~\bibnamefont
  {Fern\'andez-M\'endez}}, \bibinfo {author} {\bibfnamefont {G.~A.~M.}\
  \bibnamefont {Marug\'an}}, \ and\ \bibinfo {author} {\bibfnamefont
  {J.}~\bibnamefont {Olmedo}},\ }\href {\doibase 10.1103/PhysRevD.90.064015}
  {\bibfield  {journal} {\bibinfo  {journal} {Phys. Rev. D}\ }\textbf {\bibinfo
  {volume} {90}},\ \bibinfo {pages} {064015} (\bibinfo {year} {2014})},\
  \Eprint {http://arxiv.org/abs/1407.0998} {arXiv:1407.0998 [gr-qc]}
  \BibitemShut {NoStop}%
\bibitem [{\citenamefont {Gomar}\ \emph {et~al.}(2015)\citenamefont {Gomar},
  \citenamefont {Mart\'in-Benito},\ and\ \citenamefont {Marug\'an}}]{hyb-inf5}%
  \BibitemOpen
  \bibfield  {author} {\bibinfo {author} {\bibfnamefont {L.~C.}\ \bibnamefont
  {Gomar}}, \bibinfo {author} {\bibfnamefont {M.}~\bibnamefont
  {Mart\'in-Benito}}, \ and\ \bibinfo {author} {\bibfnamefont {G.~A.~M.}\
  \bibnamefont {Marug\'an}},\ }\href {\doibase 10.1088/1475-7516/2015/06/045}
  {\bibfield  {journal} {\bibinfo  {journal} {JCAP}\ }\textbf {\bibinfo
  {volume} {1506}},\ \bibinfo {pages} {045} (\bibinfo {year} {2015})},\ \Eprint
  {http://arxiv.org/abs/1503.03907} {arXiv:1503.03907 [gr-qc]} \BibitemShut
  {NoStop}%
\bibitem [{\citenamefont {de~Blas}\ and\ \citenamefont
  {Olmedo}(2016)}]{hyb-inf6}%
  \BibitemOpen
  \bibfield  {author} {\bibinfo {author} {\bibfnamefont {D.~M.}\ \bibnamefont
  {de~Blas}}\ and\ \bibinfo {author} {\bibfnamefont {J.}~\bibnamefont
  {Olmedo}},\ }\href {\doibase 10.1088/1475-7516/2016/06/029} {\bibfield
  {journal} {\bibinfo  {journal} {JCAP}\ }\textbf {\bibinfo {volume} {1606}},\
  \bibinfo {pages} {029} (\bibinfo {year} {2016})},\ \Eprint
  {http://arxiv.org/abs/1601.01716} {arXiv:1601.01716 [gr-qc]} \BibitemShut
  {NoStop}%
\bibitem [{\citenamefont {Martin-Benito}\ \emph {et~al.}(2011)\citenamefont
  {Martin-Benito}, \citenamefont {Martin-de Blas},\ and\ \citenamefont
  {Mena~Marugan}}]{hybrid3}%
  \BibitemOpen
  \bibfield  {author} {\bibinfo {author} {\bibfnamefont {M.}~\bibnamefont
  {Martin-Benito}}, \bibinfo {author} {\bibfnamefont {D.}~\bibnamefont
  {Martin-de Blas}}, \ and\ \bibinfo {author} {\bibfnamefont {G.~A.}\
  \bibnamefont {Mena~Marugan}},\ }\href {\doibase 10.1103/PhysRevD.83.084050}
  {\bibfield  {journal} {\bibinfo  {journal} {Phys. Rev. D}\ }\textbf {\bibinfo
  {volume} {83}},\ \bibinfo {pages} {084050} (\bibinfo {year} {2011})},\
  \Eprint {http://arxiv.org/abs/1012.2324} {arXiv:1012.2324 [gr-qc]}
  \BibitemShut {NoStop}%
\bibitem [{\citenamefont {Mart\'in-Benito}\ \emph {et~al.}(2014)\citenamefont
  {Mart\'in-Benito}, \citenamefont {Mart\'in-de Blas},\ and\ \citenamefont
  {Mena~Marug\'an}}]{hybrid4}%
  \BibitemOpen
  \bibfield  {author} {\bibinfo {author} {\bibfnamefont {M.}~\bibnamefont
  {Mart\'in-Benito}}, \bibinfo {author} {\bibfnamefont {D.}~\bibnamefont
  {Mart\'in-de Blas}}, \ and\ \bibinfo {author} {\bibfnamefont {G.~A.}\
  \bibnamefont {Mena~Marug\'an}},\ }\href {\doibase
  10.1088/0264-9381/31/7/075022} {\bibfield  {journal} {\bibinfo  {journal}
  {Class. Quant. Grav.}\ }\textbf {\bibinfo {volume} {31}},\ \bibinfo {pages}
  {075022} (\bibinfo {year} {2014})},\ \Eprint {http://arxiv.org/abs/1307.1420}
  {arXiv:1307.1420 [gr-qc]} \BibitemShut {NoStop}%
\bibitem [{\citenamefont {Elizaga~Navascu\'es}\ \emph
  {et~al.}(2015{\natexlab{a}})\citenamefont {Elizaga~Navascu\'es},
  \citenamefont {Mart\'in-Benito},\ and\ \citenamefont
  {Mena~Marug\'an}}]{hybrid5}%
  \BibitemOpen
  \bibfield  {author} {\bibinfo {author} {\bibfnamefont {B.}~\bibnamefont
  {Elizaga~Navascu\'es}}, \bibinfo {author} {\bibfnamefont {M.}~\bibnamefont
  {Mart\'in-Benito}}, \ and\ \bibinfo {author} {\bibfnamefont {G.~A.}\
  \bibnamefont {Mena~Marug\'an}},\ }\href {\doibase 10.1103/PhysRevD.91.024028}
  {\bibfield  {journal} {\bibinfo  {journal} {Phys. Rev. D}\ }\textbf {\bibinfo
  {volume} {91}},\ \bibinfo {pages} {024028} (\bibinfo {year}
  {2015}{\natexlab{a}})},\ \Eprint {http://arxiv.org/abs/1409.2927}
  {arXiv:1409.2927 [gr-qc]} \BibitemShut {NoStop}%
\bibitem [{\citenamefont {Elizaga~Navascu\'es}\ \emph
  {et~al.}(2015{\natexlab{b}})\citenamefont {Elizaga~Navascu\'es},
  \citenamefont {Mart\'in-Benito},\ and\ \citenamefont
  {Mena~Marug\'an}}]{hybrid6}%
  \BibitemOpen
  \bibfield  {author} {\bibinfo {author} {\bibfnamefont {B.}~\bibnamefont
  {Elizaga~Navascu\'es}}, \bibinfo {author} {\bibfnamefont {M.}~\bibnamefont
  {Mart\'in-Benito}}, \ and\ \bibinfo {author} {\bibfnamefont {G.~A.}\
  \bibnamefont {Mena~Marug\'an}},\ }\href {\doibase 10.1103/PhysRevD.92.024007}
  {\bibfield  {journal} {\bibinfo  {journal} {Phys. Rev. D}\ }\textbf {\bibinfo
  {volume} {92}},\ \bibinfo {pages} {024007} (\bibinfo {year}
  {2015}{\natexlab{b}})},\ \Eprint {http://arxiv.org/abs/1504.08152}
  {arXiv:1504.08152 [gr-qc]} \BibitemShut {NoStop}%
\bibitem [{\citenamefont {de~Blas}\ \emph {et~al.}(2015)\citenamefont
  {de~Blas}, \citenamefont {Olmedo},\ and\ \citenamefont
  {Pawłowski}}]{mop-lrs}%
  \BibitemOpen
  \bibfield  {author} {\bibinfo {author} {\bibfnamefont {D.~M.}\ \bibnamefont
  {de~Blas}}, \bibinfo {author} {\bibfnamefont {J.}~\bibnamefont {Olmedo}}, \
  and\ \bibinfo {author} {\bibfnamefont {T.}~\bibnamefont {Pawłowski}},\
  }\href@noop {} {\  (\bibinfo {year} {2015})},\ \Eprint
  {http://arxiv.org/abs/1509.09197} {arXiv:1509.09197 [gr-qc]} \BibitemShut
  {NoStop}%
\bibitem [{\citenamefont {Banerjee}\ \emph {et~al.}(2012)\citenamefont
  {Banerjee}, \citenamefont {Calcagni},\ and\ \citenamefont
  {Martin-Benito}}]{lqc}%
  \BibitemOpen
  \bibfield  {author} {\bibinfo {author} {\bibfnamefont {K.}~\bibnamefont
  {Banerjee}}, \bibinfo {author} {\bibfnamefont {G.}~\bibnamefont {Calcagni}},
  \ and\ \bibinfo {author} {\bibfnamefont {M.}~\bibnamefont {Martin-Benito}},\
  }\href {\doibase 10.3842/SIGMA.2012.016} {\bibfield  {journal} {\bibinfo
  {journal} {SIGMA}\ }\textbf {\bibinfo {volume} {8}},\ \bibinfo {pages} {016}
  (\bibinfo {year} {2012})},\ \Eprint {http://arxiv.org/abs/1109.6801}
  {arXiv:1109.6801 [gr-qc]} \BibitemShut {NoStop}%
\bibitem [{\citenamefont {Ashtekar}\ and\ \citenamefont
  {Wilson-Ewing}(2009)}]{awe-b1}%
  \BibitemOpen
  \bibfield  {author} {\bibinfo {author} {\bibfnamefont {A.}~\bibnamefont
  {Ashtekar}}\ and\ \bibinfo {author} {\bibfnamefont {E.}~\bibnamefont
  {Wilson-Ewing}},\ }\href {\doibase 10.1103/PhysRevD.79.083535} {\bibfield
  {journal} {\bibinfo  {journal} {Phys. Rev. D}\ }\textbf {\bibinfo {volume}
  {79}},\ \bibinfo {pages} {083535} (\bibinfo {year} {2009})},\ \Eprint
  {http://arxiv.org/abs/0903.3397} {arXiv:0903.3397 [gr-qc]} \BibitemShut
  {NoStop}%
\bibitem [{\citenamefont {Martin-Benito}\ \emph
  {et~al.}(2009{\natexlab{a}})\citenamefont {Martin-Benito}, \citenamefont
  {Marugan},\ and\ \citenamefont {Pawlowski}}]{mmp-B1-evo}%
  \BibitemOpen
  \bibfield  {author} {\bibinfo {author} {\bibfnamefont {M.}~\bibnamefont
  {Martin-Benito}}, \bibinfo {author} {\bibfnamefont {G.~A.~M.}\ \bibnamefont
  {Marugan}}, \ and\ \bibinfo {author} {\bibfnamefont {T.}~\bibnamefont
  {Pawlowski}},\ }\href {\doibase 10.1103/PhysRevD.80.084038} {\bibfield
  {journal} {\bibinfo  {journal} {Phys. Rev. D}\ }\textbf {\bibinfo {volume}
  {80}},\ \bibinfo {pages} {084038} (\bibinfo {year} {2009}{\natexlab{a}})},\
  \Eprint {http://arxiv.org/abs/0906.3751} {arXiv:0906.3751 [gr-qc]}
  \BibitemShut {NoStop}%
\bibitem [{\citenamefont {Bojowald}\ and\ \citenamefont
  {Brahma}(2015{\natexlab{a}})}]{bojo-bra}%
  \BibitemOpen
  \bibfield  {author} {\bibinfo {author} {\bibfnamefont {M.}~\bibnamefont
  {Bojowald}}\ and\ \bibinfo {author} {\bibfnamefont {S.}~\bibnamefont
  {Brahma}},\ }\href@noop {} {\bibfield  {journal} {\bibinfo  {journal} {Phys.
  Rev. D}\ }\textbf {\bibinfo {volume} {92}},\ \bibinfo {pages} {065002}
  (\bibinfo {year} {2015}{\natexlab{a}})}\BibitemShut {NoStop}%
\bibitem [{\citenamefont {Gambini}\ and\ \citenamefont {Pullin}(2013)}]{BH-1}%
  \BibitemOpen
  \bibfield  {author} {\bibinfo {author} {\bibfnamefont {R.}~\bibnamefont
  {Gambini}}\ and\ \bibinfo {author} {\bibfnamefont {J.}~\bibnamefont
  {Pullin}},\ }\href {\doibase 10.1103/PhysRevLett.110.211301} {\bibfield
  {journal} {\bibinfo  {journal} {Phys. Rev. Lett.}\ }\textbf {\bibinfo
  {volume} {110}},\ \bibinfo {pages} {211301} (\bibinfo {year} {2013})},\
  \Eprint {http://arxiv.org/abs/1302.5265} {arXiv:1302.5265 [gr-qc]}
  \BibitemShut {NoStop}%
\bibitem [{\citenamefont {Gambini}\ \emph {et~al.}(2014)\citenamefont
  {Gambini}, \citenamefont {Olmedo},\ and\ \citenamefont {Pullin}}]{BH-2}%
  \BibitemOpen
  \bibfield  {author} {\bibinfo {author} {\bibfnamefont {R.}~\bibnamefont
  {Gambini}}, \bibinfo {author} {\bibfnamefont {J.}~\bibnamefont {Olmedo}}, \
  and\ \bibinfo {author} {\bibfnamefont {J.}~\bibnamefont {Pullin}},\ }\href
  {\doibase 10.1088/0264-9381/31/9/095009} {\bibfield  {journal} {\bibinfo
  {journal} {Class. Quant. Grav.}\ }\textbf {\bibinfo {volume} {31}},\ \bibinfo
  {pages} {095009} (\bibinfo {year} {2014})},\ \Eprint
  {http://arxiv.org/abs/1310.5996} {arXiv:1310.5996 [gr-qc]} \BibitemShut
  {NoStop}%
\bibitem [{\citenamefont {Bojowald}\ and\ \citenamefont
  {Brahma}(2015{\natexlab{b}})}]{bb-gowdy}%
  \BibitemOpen
  \bibfield  {author} {\bibinfo {author} {\bibfnamefont {M.}~\bibnamefont
  {Bojowald}}\ and\ \bibinfo {author} {\bibfnamefont {S.}~\bibnamefont
  {Brahma}},\ }\href {\doibase 10.1103/PhysRevD.92.065002} {\bibfield
  {journal} {\bibinfo  {journal} {Phys. Rev. D}\ }\textbf {\bibinfo {volume}
  {92}},\ \bibinfo {pages} {065002} (\bibinfo {year} {2015}{\natexlab{b}})},\
  \Eprint {http://arxiv.org/abs/1507.00679} {arXiv:1507.00679 [gr-qc]}
  \BibitemShut {NoStop}%
\bibitem [{\citenamefont {Immirzi}(1997)}]{immirzi}%
  \BibitemOpen
  \bibfield  {author} {\bibinfo {author} {\bibfnamefont {G.}~\bibnamefont
  {Immirzi}},\ }\href {\doibase 10.1088/0264-9381/14/10/002} {\bibfield
  {journal} {\bibinfo  {journal} {Class. Quant. Grav.}\ }\textbf {\bibinfo
  {volume} {14}},\ \bibinfo {pages} {L177} (\bibinfo {year} {1997})},\ \Eprint
  {http://arxiv.org/abs/gr-qc/9612030} {arXiv:gr-qc/9612030 [gr-qc]}
  \BibitemShut {NoStop}%
\bibitem [{\citenamefont {Hajicek}\ and\ \citenamefont
  {Kuchar}(1990)}]{kuchar}%
  \BibitemOpen
  \bibfield  {author} {\bibinfo {author} {\bibfnamefont {P.}~\bibnamefont
  {Hajicek}}\ and\ \bibinfo {author} {\bibfnamefont {K.~V.}\ \bibnamefont
  {Kuchar}},\ }\href {\doibase 10.1103/PhysRevD.41.1091} {\bibfield  {journal}
  {\bibinfo  {journal} {Phys. Rev. D}\ }\textbf {\bibinfo {volume} {41}},\
  \bibinfo {pages} {1091} (\bibinfo {year} {1990})}\BibitemShut {NoStop}%
\bibitem [{\citenamefont {Henneaux}\ and\ \citenamefont
  {Teitelboim}(1992)}]{teitel}%
  \BibitemOpen
  \bibfield  {author} {\bibinfo {author} {\bibfnamefont {M.}~\bibnamefont
  {Henneaux}}\ and\ \bibinfo {author} {\bibfnamefont {C.}~\bibnamefont
  {Teitelboim}},\ }\href@noop {} {\emph {\bibinfo {title} {{Quantization of
  gauge systems}}}}\ (\bibinfo {year} {1992})\BibitemShut {NoStop}%
\bibitem [{\citenamefont {Bojowald}(2004)}]{b-spher}%
  \BibitemOpen
  \bibfield  {author} {\bibinfo {author} {\bibfnamefont {M.}~\bibnamefont
  {Bojowald}},\ }\href {\doibase 10.1088/0264-9381/21/15/008} {\bibfield
  {journal} {\bibinfo  {journal} {Class. Quant. Grav.}\ }\textbf {\bibinfo
  {volume} {21}},\ \bibinfo {pages} {3733} (\bibinfo {year} {2004})},\ \Eprint
  {http://arxiv.org/abs/gr-qc/0407017} {arXiv:gr-qc/0407017 [gr-qc]}
  \BibitemShut {NoStop}%
\bibitem [{\citenamefont {Bojowald}\ and\ \citenamefont
  {Swiderski}(2006)}]{bs-ham}%
  \BibitemOpen
  \bibfield  {author} {\bibinfo {author} {\bibfnamefont {M.}~\bibnamefont
  {Bojowald}}\ and\ \bibinfo {author} {\bibfnamefont {R.}~\bibnamefont
  {Swiderski}},\ }\href {\doibase 10.1088/0264-9381/23/6/015} {\bibfield
  {journal} {\bibinfo  {journal} {Class. Quant. Grav.}\ }\textbf {\bibinfo
  {volume} {23}},\ \bibinfo {pages} {2129} (\bibinfo {year} {2006})},\ \Eprint
  {http://arxiv.org/abs/gr-qc/0511108} {arXiv:gr-qc/0511108 [gr-qc]}
  \BibitemShut {NoStop}%
\bibitem [{\citenamefont {Ashtekar}\ \emph {et~al.}(2006)\citenamefont
  {Ashtekar}, \citenamefont {Pawlowski},\ and\ \citenamefont {Singh}}]{aps}%
  \BibitemOpen
  \bibfield  {author} {\bibinfo {author} {\bibfnamefont {A.}~\bibnamefont
  {Ashtekar}}, \bibinfo {author} {\bibfnamefont {T.}~\bibnamefont {Pawlowski}},
  \ and\ \bibinfo {author} {\bibfnamefont {P.}~\bibnamefont {Singh}},\ }\href
  {\doibase 10.1103/PhysRevD.74.084003} {\bibfield  {journal} {\bibinfo
  {journal} {Phys. Rev. D}\ }\textbf {\bibinfo {volume} {74}},\ \bibinfo
  {pages} {084003} (\bibinfo {year} {2006})},\ \Eprint
  {http://arxiv.org/abs/gr-qc/0607039} {arXiv:gr-qc/0607039 [gr-qc]}
  \BibitemShut {NoStop}%
\bibitem [{\citenamefont {Chiou}\ \emph {et~al.}(2012)\citenamefont {Chiou},
  \citenamefont {Ni},\ and\ \citenamefont {Tang}}]{chiouBH}%
  \BibitemOpen
  \bibfield  {author} {\bibinfo {author} {\bibfnamefont {D.-W.}\ \bibnamefont
  {Chiou}}, \bibinfo {author} {\bibfnamefont {W.-T.}\ \bibnamefont {Ni}}, \
  and\ \bibinfo {author} {\bibfnamefont {A.}~\bibnamefont {Tang}},\ }\href@noop
  {} {\  (\bibinfo {year} {2012})},\ \Eprint {http://arxiv.org/abs/1212.1265}
  {arXiv:1212.1265 [gr-qc]} \BibitemShut {NoStop}%
\bibitem [{\citenamefont {Singh}\ and\ \citenamefont
  {Vandersloot}(2005)}]{sv-eff}%
  \BibitemOpen
  \bibfield  {author} {\bibinfo {author} {\bibfnamefont {P.}~\bibnamefont
  {Singh}}\ and\ \bibinfo {author} {\bibfnamefont {K.}~\bibnamefont
  {Vandersloot}},\ }\href {\doibase 10.1103/PhysRevD.72.084004} {\bibfield
  {journal} {\bibinfo  {journal} {Phys. Rev. D}\ }\textbf {\bibinfo {volume}
  {72}},\ \bibinfo {pages} {084004} (\bibinfo {year} {2005})},\ \Eprint
  {http://arxiv.org/abs/gr-qc/0507029} {arXiv:gr-qc/0507029 [gr-qc]}
  \BibitemShut {NoStop}%
\bibitem [{\citenamefont {Thiemann}(1998)}]{t-qsd5}%
  \BibitemOpen
  \bibfield  {author} {\bibinfo {author} {\bibfnamefont {T.}~\bibnamefont
  {Thiemann}},\ }\href {\doibase 10.1088/0264-9381/15/5/012} {\bibfield
  {journal} {\bibinfo  {journal} {Class. Quant. Grav.}\ }\textbf {\bibinfo
  {volume} {15}},\ \bibinfo {pages} {1281} (\bibinfo {year} {1998})},\ \Eprint
  {http://arxiv.org/abs/gr-qc/9705019} {arXiv:gr-qc/9705019 [gr-qc]}
  \BibitemShut {NoStop}%
\bibitem [{\citenamefont {Ashtekar}\ \emph {et~al.}(2003)\citenamefont
  {Ashtekar}, \citenamefont {Bojowald},\ and\ \citenamefont
  {Lewandowski}}]{abl-lqc}%
  \BibitemOpen
  \bibfield  {author} {\bibinfo {author} {\bibfnamefont {A.}~\bibnamefont
  {Ashtekar}}, \bibinfo {author} {\bibfnamefont {M.}~\bibnamefont {Bojowald}},
  \ and\ \bibinfo {author} {\bibfnamefont {J.}~\bibnamefont {Lewandowski}},\
  }\href {\doibase 10.4310/ATMP.2003.v7.n2.a2} {\bibfield  {journal} {\bibinfo
  {journal} {Adv. Theor. Math. Phys.}\ }\textbf {\bibinfo {volume} {7}},\
  \bibinfo {pages} {233} (\bibinfo {year} {2003})},\ \Eprint
  {http://arxiv.org/abs/gr-qc/0304074} {arXiv:gr-qc/0304074 [gr-qc]}
  \BibitemShut {NoStop}%
\bibitem [{\citenamefont {Ashtekar}\ and\ \citenamefont
  {Lewandowski}(1998)}]{al-vol}%
  \BibitemOpen
  \bibfield  {author} {\bibinfo {author} {\bibfnamefont {A.}~\bibnamefont
  {Ashtekar}}\ and\ \bibinfo {author} {\bibfnamefont {J.}~\bibnamefont
  {Lewandowski}},\ }\href@noop {} {\bibfield  {journal} {\bibinfo  {journal}
  {Adv. Theor. Math. Phys.}\ }\textbf {\bibinfo {volume} {1}},\ \bibinfo
  {pages} {388} (\bibinfo {year} {1998})},\ \Eprint
  {http://arxiv.org/abs/gr-qc/9711031} {arXiv:gr-qc/9711031 [gr-qc]}
  \BibitemShut {NoStop}%
\bibitem [{\citenamefont {Bojowald}(2000)}]{b-vol}%
  \BibitemOpen
  \bibfield  {author} {\bibinfo {author} {\bibfnamefont {M.}~\bibnamefont
  {Bojowald}},\ }\href {\doibase 10.1088/0264-9381/17/6/313} {\bibfield
  {journal} {\bibinfo  {journal} {Class. Quant. Grav.}\ }\textbf {\bibinfo
  {volume} {17}},\ \bibinfo {pages} {1509} (\bibinfo {year} {2000})},\ \Eprint
  {http://arxiv.org/abs/gr-qc/9910104} {arXiv:gr-qc/9910104 [gr-qc]}
  \BibitemShut {NoStop}%
\bibitem [{\citenamefont {Martin-Benito}\ \emph
  {et~al.}(2008{\natexlab{b}})\citenamefont {Martin-Benito}, \citenamefont
  {Mena~Marugan},\ and\ \citenamefont {Pawlowski}}]{mmp}%
  \BibitemOpen
  \bibfield  {author} {\bibinfo {author} {\bibfnamefont {M.}~\bibnamefont
  {Martin-Benito}}, \bibinfo {author} {\bibfnamefont {G.~A.}\ \bibnamefont
  {Mena~Marugan}}, \ and\ \bibinfo {author} {\bibfnamefont {T.}~\bibnamefont
  {Pawlowski}},\ }\href {\doibase 10.1103/PhysRevD.78.064008} {\bibfield
  {journal} {\bibinfo  {journal} {Phys. Rev. D}\ }\textbf {\bibinfo {volume}
  {78}},\ \bibinfo {pages} {064008} (\bibinfo {year} {2008}{\natexlab{b}})},\
  \Eprint {http://arxiv.org/abs/0804.3157} {arXiv:0804.3157 [gr-qc]}
  \BibitemShut {NoStop}%
\bibitem [{\citenamefont {Martin-Benito}\ \emph
  {et~al.}(2009{\natexlab{b}})\citenamefont {Martin-Benito}, \citenamefont
  {Marugan},\ and\ \citenamefont {Olmedo}}]{mmo-frw}%
  \BibitemOpen
  \bibfield  {author} {\bibinfo {author} {\bibfnamefont {M.}~\bibnamefont
  {Martin-Benito}}, \bibinfo {author} {\bibfnamefont {G.~A.~M.}\ \bibnamefont
  {Marugan}}, \ and\ \bibinfo {author} {\bibfnamefont {J.}~\bibnamefont
  {Olmedo}},\ }\href {\doibase 10.1103/PhysRevD.80.104015} {\bibfield
  {journal} {\bibinfo  {journal} {Phys. Rev. D}\ }\textbf {\bibinfo {volume}
  {80}},\ \bibinfo {pages} {104015} (\bibinfo {year} {2009}{\natexlab{b}})},\
  \Eprint {http://arxiv.org/abs/0909.2829} {arXiv:0909.2829 [gr-qc]}
  \BibitemShut {NoStop}%
\bibitem [{\citenamefont {Mena~Marugan}\ \emph {et~al.}(2011)\citenamefont
  {Mena~Marugan}, \citenamefont {Olmedo},\ and\ \citenamefont
  {Pawlowski}}]{mmp-presc}%
  \BibitemOpen
  \bibfield  {author} {\bibinfo {author} {\bibfnamefont {G.~A.}\ \bibnamefont
  {Mena~Marugan}}, \bibinfo {author} {\bibfnamefont {J.}~\bibnamefont
  {Olmedo}}, \ and\ \bibinfo {author} {\bibfnamefont {T.}~\bibnamefont
  {Pawlowski}},\ }\href {\doibase 10.1103/PhysRevD.84.064012} {\bibfield
  {journal} {\bibinfo  {journal} {Phys. Rev. D}\ }\textbf {\bibinfo {volume}
  {84}},\ \bibinfo {pages} {064012} (\bibinfo {year} {2011})},\ \Eprint
  {http://arxiv.org/abs/1108.0829} {arXiv:1108.0829 [gr-qc]} \BibitemShut
  {NoStop}%
\bibitem [{\citenamefont {Berger}(2002)}]{Berger:2002st}%
  \BibitemOpen
  \bibfield  {author} {\bibinfo {author} {\bibfnamefont {B.~K.}\ \bibnamefont
  {Berger}},\ }\href {\doibase 10.12942/lrr-2002-1} {\bibfield  {journal}
  {\bibinfo  {journal} {Living Rev. Rel.}\ }\textbf {\bibinfo {volume} {5}},\
  \bibinfo {pages} {1} (\bibinfo {year} {2002})},\ \Eprint
  {http://arxiv.org/abs/gr-qc/0201056} {arXiv:gr-qc/0201056 [gr-qc]}
  \BibitemShut {NoStop}%
\bibitem [{\citenamefont {Laddha}\ and\ \citenamefont
  {Varadarajan}(2011)}]{mad}%
  \BibitemOpen
  \bibfield  {author} {\bibinfo {author} {\bibfnamefont {A.}~\bibnamefont
  {Laddha}}\ and\ \bibinfo {author} {\bibfnamefont {M.}~\bibnamefont
  {Varadarajan}},\ }\href {\doibase 10.1088/0264-9381/28/19/195010} {\bibfield
  {journal} {\bibinfo  {journal} {Class. Quant. Grav.}\ }\textbf {\bibinfo
  {volume} {28}},\ \bibinfo {pages} {195010} (\bibinfo {year} {2011})},\
  \Eprint {http://arxiv.org/abs/1105.0636} {arXiv:1105.0636 [gr-qc]}
  \BibitemShut {NoStop}%
\bibitem [{\citenamefont {Ashtekar}\ \emph {et~al.}(1995)\citenamefont
  {Ashtekar}, \citenamefont {Lewandowski}, \citenamefont {Marolf},
  \citenamefont {Mourao},\ and\ \citenamefont {Thiemann}}]{ga}%
  \BibitemOpen
  \bibfield  {author} {\bibinfo {author} {\bibfnamefont {A.}~\bibnamefont
  {Ashtekar}}, \bibinfo {author} {\bibfnamefont {J.}~\bibnamefont
  {Lewandowski}}, \bibinfo {author} {\bibfnamefont {D.}~\bibnamefont {Marolf}},
  \bibinfo {author} {\bibfnamefont {J.}~\bibnamefont {Mourao}}, \ and\ \bibinfo
  {author} {\bibfnamefont {T.}~\bibnamefont {Thiemann}},\ }\href {\doibase
  10.1063/1.531252} {\bibfield  {journal} {\bibinfo  {journal} {J. Math.
  Phys.}\ }\textbf {\bibinfo {volume} {36}},\ \bibinfo {pages} {6456} (\bibinfo
  {year} {1995})},\ \Eprint {http://arxiv.org/abs/gr-qc/9504018}
  {arXiv:gr-qc/9504018 [gr-qc]} \BibitemShut {NoStop}%
\bibitem [{\citenamefont {Marolf}(1995{\natexlab{a}})}]{m-gave1}%
  \BibitemOpen
  \bibfield  {author} {\bibinfo {author} {\bibfnamefont {D.}~\bibnamefont
  {Marolf}},\ }\href@noop {} {\  (\bibinfo {year} {1995}{\natexlab{a}})},\
  \Eprint {http://arxiv.org/abs/gr-qc/9508015} {arXiv:gr-qc/9508015}
  \BibitemShut {NoStop}%
\bibitem [{\citenamefont {Marolf}(1995{\natexlab{b}})}]{m-gave2}%
  \BibitemOpen
  \bibfield  {author} {\bibinfo {author} {\bibfnamefont {D.}~\bibnamefont
  {Marolf}},\ }\href@noop {} {\bibfield  {journal} {\bibinfo  {journal} {Class.
  Quant. Grav.}\ }\textbf {\bibinfo {volume} {12}},\ \bibinfo {pages} {1199}
  (\bibinfo {year} {1995}{\natexlab{b}})},\ \Eprint
  {http://arxiv.org/abs/gr-qc/9404053} {arXiv:gr-qc/9404053} \BibitemShut
  {NoStop}%
\bibitem [{\citenamefont {Marolf}(1995{\natexlab{c}})}]{m-gave3}%
  \BibitemOpen
  \bibfield  {author} {\bibinfo {author} {\bibfnamefont {D.}~\bibnamefont
  {Marolf}},\ }\href@noop {} {\bibfield  {journal} {\bibinfo  {journal} {Class.
  Quant. Grav.}\ }\textbf {\bibinfo {volume} {12}},\ \bibinfo {pages} {1441}
  (\bibinfo {year} {1995}{\natexlab{c}})},\ \Eprint
  {http://arxiv.org/abs/gr-qc/9409049} {arXiv:gr-qc/9409049} \BibitemShut
  {NoStop}%
\bibitem [{\citenamefont {Marolf}(1995{\natexlab{d}})}]{m-gave4}%
  \BibitemOpen
  \bibfield  {author} {\bibinfo {author} {\bibfnamefont {D.}~\bibnamefont
  {Marolf}},\ }\href {\doibase 10.1088/0264-9381/12/10/007} {\bibfield
  {journal} {\bibinfo  {journal} {Class. Quant. Grav.}\ }\textbf {\bibinfo
  {volume} {12}},\ \bibinfo {pages} {2469} (\bibinfo {year}
  {1995}{\natexlab{d}})},\ \Eprint {http://arxiv.org/abs/gr-qc/9412016}
  {arXiv:gr-qc/9412016} \BibitemShut {NoStop}%
\bibitem [{\citenamefont {Rovelli}(1990)}]{carlo0}%
  \BibitemOpen
  \bibfield  {author} {\bibinfo {author} {\bibfnamefont {C.}~\bibnamefont
  {Rovelli}},\ }\href@noop {} {\bibfield  {journal} {\bibinfo  {journal} {Phys.
  Rev. D}\ }\textbf {\bibinfo {volume} {42}},\ \bibinfo {pages} {2638}
  (\bibinfo {year} {1990})}\BibitemShut {NoStop}%
\bibitem [{\citenamefont {Rovelli}(1991)}]{carlo1}%
  \BibitemOpen
  \bibfield  {author} {\bibinfo {author} {\bibfnamefont {C.}~\bibnamefont
  {Rovelli}},\ }\href@noop {} {\bibfield  {journal} {\bibinfo  {journal} {Phys.
  Rev. D}\ }\textbf {\bibinfo {volume} {43}},\ \bibinfo {pages} {442} (\bibinfo
  {year} {1991})}\BibitemShut {NoStop}%
\bibitem [{\citenamefont {Dittrich}(2006)}]{bianca0}%
  \BibitemOpen
  \bibfield  {author} {\bibinfo {author} {\bibfnamefont {B.}~\bibnamefont
  {Dittrich}},\ }\href@noop {} {\bibfield  {journal} {\bibinfo  {journal}
  {Class. Quant. Grav.}\ }\textbf {\bibinfo {volume} {23}},\ \bibinfo {pages}
  {6155} (\bibinfo {year} {2006})}\BibitemShut {NoStop}%
\bibitem [{\citenamefont {Dittrich}(2007)}]{bianca1}%
  \BibitemOpen
  \bibfield  {author} {\bibinfo {author} {\bibfnamefont {B.}~\bibnamefont
  {Dittrich}},\ }\href@noop {} {\bibfield  {journal} {\bibinfo  {journal} {Gen.
  Rel. Grav.}\ }\textbf {\bibinfo {volume} {39}},\ \bibinfo {pages} {1891}
  (\bibinfo {year} {2007})}\BibitemShut {NoStop}%
\bibitem [{\citenamefont {Kaminski}\ and\ \citenamefont
  {Pawlowski}(2010)}]{kp-scatter}%
  \BibitemOpen
  \bibfield  {author} {\bibinfo {author} {\bibfnamefont {W.}~\bibnamefont
  {Kaminski}}\ and\ \bibinfo {author} {\bibfnamefont {T.}~\bibnamefont
  {Pawlowski}},\ }\href {\doibase 10.1103/PhysRevD.81.084027} {\bibfield
  {journal} {\bibinfo  {journal} {Phys. Rev. D}\ }\textbf {\bibinfo {volume}
  {81}},\ \bibinfo {pages} {084027} (\bibinfo {year} {2010})},\ \Eprint
  {http://arxiv.org/abs/1001.2663} {arXiv:1001.2663 [gr-qc]} \BibitemShut
  {NoStop}%
\bibitem [{\citenamefont {Pawlowski}\ \emph {et~al.}(2014)\citenamefont
  {Pawlowski}, \citenamefont {Pierini},\ and\ \citenamefont
  {Wilson-Ewing}}]{ppwe-rad}%
  \BibitemOpen
  \bibfield  {author} {\bibinfo {author} {\bibfnamefont {T.}~\bibnamefont
  {Pawlowski}}, \bibinfo {author} {\bibfnamefont {R.}~\bibnamefont {Pierini}},
  \ and\ \bibinfo {author} {\bibfnamefont {E.}~\bibnamefont {Wilson-Ewing}},\
  }\href {\doibase 10.1103/PhysRevD.90.123538} {\bibfield  {journal} {\bibinfo
  {journal} {Phys. Rev. D}\ }\textbf {\bibinfo {volume} {90}},\ \bibinfo
  {pages} {123538} (\bibinfo {year} {2014})},\ \Eprint
  {http://arxiv.org/abs/1404.4036} {arXiv:1404.4036 [gr-qc]} \BibitemShut
  {NoStop}%
\bibitem [{\citenamefont {Pawłowski}(2016)}]{p-coh}%
  \BibitemOpen
  \bibfield  {author} {\bibinfo {author} {\bibfnamefont {T.}~\bibnamefont
  {Pawłowski}},\ }\href {\doibase 10.1142/S021827181642013X} {\bibfield
  {journal} {\bibinfo  {journal} {Int. J. Mod. Phys. D}\ }\textbf {\bibinfo
  {volume} {25}},\ \bibinfo {pages} {1642013} (\bibinfo {year} {2016})},\
  \Eprint {http://arxiv.org/abs/1607.04121} {arXiv:1607.04121 [gr-qc]}
  \BibitemShut {NoStop}%
\bibitem [{\citenamefont {Bojowald}\ \emph {et~al.}(2015)\citenamefont
  {Bojowald}, \citenamefont {Brahma},\ and\ \citenamefont
  {Reyes}}]{bojo-bra-rey}%
  \BibitemOpen
  \bibfield  {author} {\bibinfo {author} {\bibfnamefont {M.}~\bibnamefont
  {Bojowald}}, \bibinfo {author} {\bibfnamefont {S.}~\bibnamefont {Brahma}}, \
  and\ \bibinfo {author} {\bibfnamefont {J.~D.}\ \bibnamefont {Reyes}},\
  }\href@noop {} {\bibfield  {journal} {\bibinfo  {journal} {Phys. Rev. D}\
  }\textbf {\bibinfo {volume} {92}},\ \bibinfo {pages} {045043} (\bibinfo
  {year} {2015})}\BibitemShut {NoStop}%
\bibitem [{\citenamefont {Campiglia}\ \emph {et~al.}(2016)\citenamefont
  {Campiglia}, \citenamefont {Gambini}, \citenamefont {Olmedo},\ and\
  \citenamefont {Pullin}}]{cgop}%
  \BibitemOpen
  \bibfield  {author} {\bibinfo {author} {\bibfnamefont {M.}~\bibnamefont
  {Campiglia}}, \bibinfo {author} {\bibfnamefont {R.}~\bibnamefont {Gambini}},
  \bibinfo {author} {\bibfnamefont {J.}~\bibnamefont {Olmedo}}, \ and\ \bibinfo
  {author} {\bibfnamefont {J.}~\bibnamefont {Pullin}},\ }\href@noop {}
  {\bibfield  {journal} {\bibinfo  {journal} {Class. Quant. Grav.}\ }\textbf
  {\bibinfo {volume} {33}},\ \bibinfo {pages} {18LT01} (\bibinfo {year}
  {2016})}\BibitemShut {NoStop}%
\end{thebibliography}%

\end{document}